\documentclass[aps,prb,twocolumn,showpacs,amsmath,amssymb,floatfix,superscriptaddress,reprint]{revtex4-2}
\usepackage[utf8]{inputenc}
\usepackage{natbib}
\usepackage[dvips]{graphics}
\usepackage[colorlinks=true,citecolor=blue]{hyperref}

\usepackage{subfigure,overpic}
\usepackage{comment}

\newcommand{\up}{\uparrow}
\newcommand{\down}{\downarrow}

\def \beq {\begin{equation}}
\def \edq {\end{equation}}
\def \bes {\begin{subequations}}
\def \eds {\end{subequations}}
\def \beqn {\begin{equation*}}
\def \edqn {\end{equation*}}

\def \dag {\dagger}

\def \up {\uparrow}
\def \down {\downarrow}

\def \veps {\varepsilon}

\def \calh {{\cal{H}}}

\def \ket {\rangle}
\def \bra {\langle}

\begin{document}
\title{Superconductor-quantum dot hybrid coolers}
\author{Sun-Yong Hwang}
\affiliation{Theoretische Physik, Universit\"at Duisburg-Essen and CENIDE, D-47048 Duisburg, Germany}
\author{Björn Sothmann}
\affiliation{Theoretische Physik, Universit\"at Duisburg-Essen and CENIDE, D-47048 Duisburg, Germany}
\author{David S\' anchez}
\affiliation{Institute for Cross-Disciplinary Physics and Complex Systems 
IFISC (UIB-CSIC), E-07122 Palma de Mallorca, Spain}
\date{\today}

\begin{abstract}
We propose a refrigeration scheme in a mesoscopic superconductor-quantum dot hybrid device. The setup can significantly cool down a normal metal coupled to the device by applying a bias voltage across the system. We demonstrate that the cooling power can be as large as 0.05$\Delta_0^2/h$ where $\Delta_0$ is the absolute value of superconducting order parameter. In contrast to previous proposals, our device operates without any magnetic elements such as ferromagnetic reservoirs or Zeeman splittings. The refrigeration scheme works over a broad parameter range and can be optimized by tuning system parameters such as level position and bias voltage. Our theory self-consistently determines the temperature drop of the normal reservoir in the nonlinear transport regime including electron-electron interactions at the mean field level. Finally, we evaluate the refrigeration performance and find efficiencies as large as half of the Carnot bound for realistic values of the coupling strength.
\end{abstract}

\maketitle
\section{Introduction}
Heat management of nanodevices is becoming of utmost importance as the increased production of heat is usually detrimental for the device functionality. One can deal with the waste heat by recovering it via thermoelectric energy harvesting~\cite{sothmann_thermoelectric_2015,benenti2017fundamental,arrachea2022energy}. Alternatively, one may achieve control over heat flows in the framework of phase-coherent caloritronics~\cite{fornieri_towards_2017, Hwang2020,PhysRevB.102.241302} where the underlying heat can be carried by phonons~\cite{li_colloquium:_2012} or electrons~\cite{giazotto_opportunities_2006}.

In order to circumvent heating and hence enhance the device performance, electrical work can be used to extract heat from the reservoir, i.e., the Peltier effect, constituting nanocoolers. There have been several proposals of such refrigerators based on two-terminal systems with resonant tunneling through quantum dots \cite{edwards_a_1993,edwards_cryogenic_1995,prance_electronic_2009,gasparinetti_probing_2011}, Coulomb-blockaded quantum dots and metallic islands \cite{timofeev_electronic_2009,arrachea_heat_2007,rey_nonadiabatic_2007,cleuren_cooling_2012,
bruggemann_cooling_2014,pekola_refrigerator_2014}, superconductor-normal metal junctions \cite{nahum_electronic_1994,leivo_efficient_1996,giazotto_opportunities_2006,muhonen_micrometre-scale_2012}, and hybrid spin-split superconductors \cite{rouco_electron_2018,bergeret_colloquium_2018}. Various multiterminal refrigerators have also been studied in the literature \cite{linden_how_2010,brunner_virtual_2012,levy_quantum_2012,brunner_entanglement_2014,correa_quantum-enhanced_2014,correa_multistage_2014,venturelli_minimal_2013,entin-wohlman_enhanced_2015,sanchez_correlation-induced_2017,erdman2018absorption,wang_nonlinear_2018,sanchez_cooling_2018,hussein_nonlocal_2019,sanchez_nonlinear_2019,dare2019comparative}.

Refrigeration schemes based on superconductor-normal metal (S-N) hybrid junctions \cite{nahum_electronic_1994,leivo_efficient_1996,tabatabaei2022nonlocal} rely on the tunneling process where the
gapped density of states in S renders a selective transport of hot carriers out of N by applying a proper bias voltage \cite{giazotto_opportunities_2006}, hence, cooling the N reservoir. While the refrigeration in this hybrid structure is usually a nonlinear effect in the applied voltage, spin-split superconductors can yield linear-in-voltage cooling effects when ferromagnetic insulating barriers are present \cite{rouco_electron_2018}.

In this paper, we consider a hybrid structure consisting of a single-level quantum dot (D) tunnel coupled to a superconductor and a normal metal reservoir. Compared to a simple tunnel junction, the considered setup offers a greater degree of tunability as the level position of the dot can be controlled by a gate voltage. This N-D-S device has been investigated in the context of thermoelectric engines \cite{hwang_large_2016} and diodes \cite{hwang_hybrid_2016} where the thermoelectric and rectification efficiency can be boosted when a magnetic field splits the quantum dot levels in a spin-dependent manner combined with the proper gate voltages. The thermoelectric efficiency can be further enhanced if the N is replaced by a ferromagnetic reservoir possessing an intrinsic spin polarization $p$. However, its potential as a refrigerator has never been explored until now. More generally, superconductor-quantum dot (S-D) hybrid devices have attracted a great deal of attention both theoretically \cite{martin-rodero_josephson_2011} and experimentally \cite{de_franceschi_hybrid_2010}. In particular, attention has been paid to Josephson effects \cite{van_dam_supercurrent_2006,jarillo-herrero_quantum_2006,jorgensen_critical_2007,baba_superconducting_2015,szombati_josephson_2016,probst_signatures_2016}, multiple Andreev reflections \cite{levy_yeyati_resonant_1997,buitelaar_multiple_2003,cuevas_full_2003,nilsson_supercurrent_2011,rentrop_nonequilibrium_2014,Hwang2017}, the Kondo effect \cite{clerk_loss_2000,buitelaar_quantum_2002,avishai_superconductor-quantum_2003,eichler_even-odd_2007,karrasch_josephson_2008}, unconventional superconducting correlations induced in quantum dots \cite{sothmann_unconventional_2014,kashuba_majorana_2017,weiss_odd-triplet_2017,hwang_odd-frequency_2018,tabatabaei2020andreev}, Cooper pair splitting \cite{recher_andreev_2001,hofstetter_cooper_2009,herrmann_carbon_2010,hofstetter_finite-bias_2011,das_high-efficiency_2012,schindele_near-unity_2012}, and the generation of Majorana fermions \cite{leijnse_parity_2012,sothmann_fractional_2013,fulga_adaptive_2013,deng_majorana_2016}.

\begin{figure}[b]
\centering
  \begin{centering}
    \includegraphics[width=0.35\textwidth,clip]{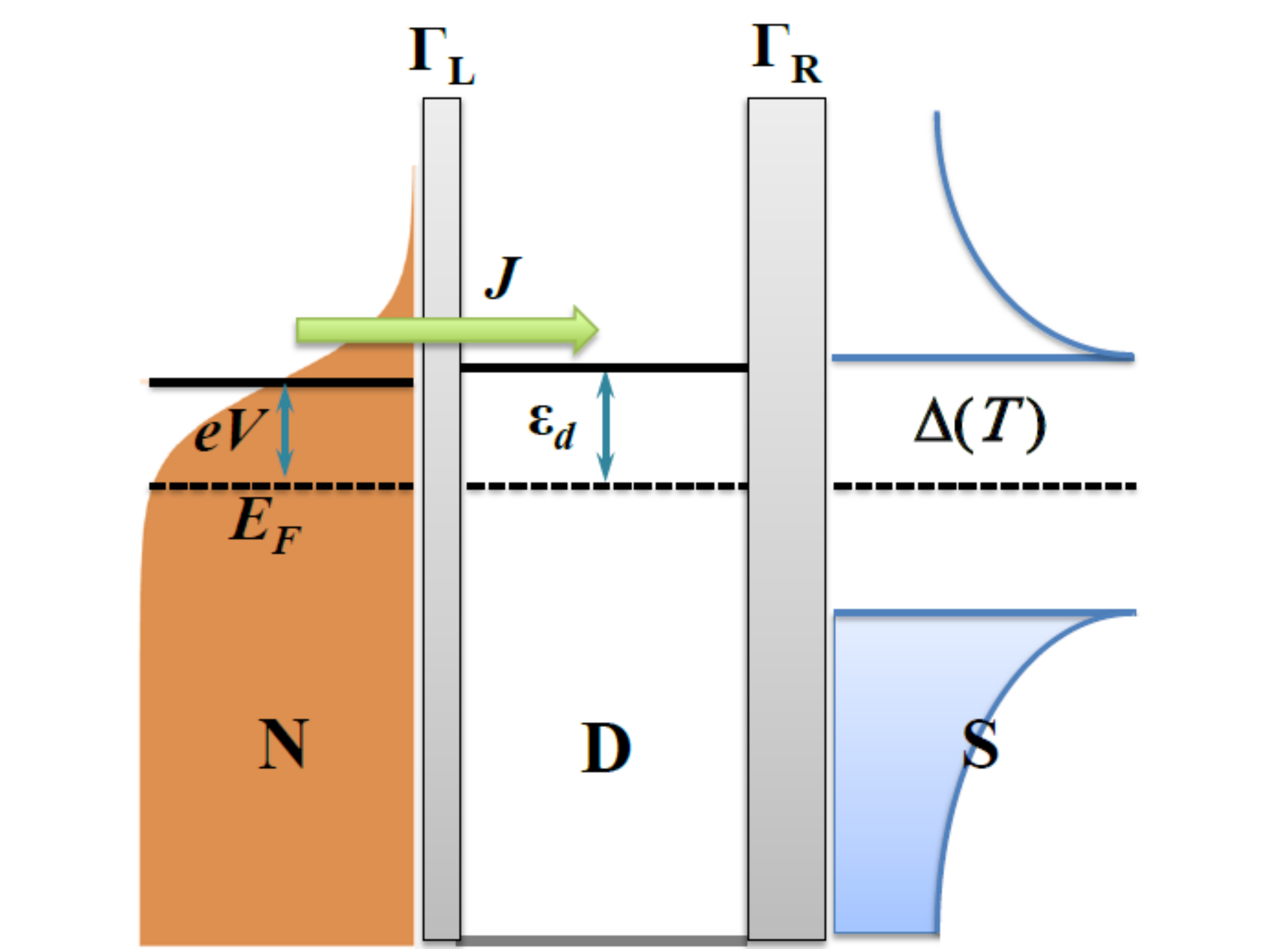}
  \end{centering}
  \caption{Energy diagram illustrating the normal metal-quantum dot-superconductor (N-D-S) refrigerator where N is to be cooled ($J>0$) via D-S hybrid structure with an optimally applied voltage $eV$ and the gate detuning of the quantum dot level $\veps_d$. The common Fermi level is denoted by the dashed lines which we take as an energy reference $E_F=0$.}
  \label{fig:sketch}
\end{figure}

The paper is organized as follows. In Sec. \ref{theory}, we discuss the theoretical methods used in this work. We denote our device as N-D-S, see Fig. \ref{fig:sketch}, and are specifically interested in cooling the N side of the device where the positive heat flow ($J>0$) corresponds to refrigeration of N with a corresponding temperature drop $\delta T<0$. Section \ref{results} is devoted to the main results of this paper followed by the summary in Sec. \ref{conclusion}. 

\section{Theoretical model}\label{theory}
Our superconductor-quantum dot refrigerator comprises the left normal metal (N) at temperature $T_N$ which is to be cooled, a single-level quantum dot (D), and the right superconductor (S) at base temperature $T$ as depicted in Fig. \ref{fig:sketch}. The total Hamiltonian reads
\beq\label{eq_calh}
\calh=\calh_{N}+\calh_{S}+\calh_{D}+\calh_{T}\,,
\edq
where 
\begin{equation}
\calh_{N}=\sum_{k\sigma}\varepsilon_{Nk\sigma}c_{Nk\sigma}^{\dag}c_{Nk\sigma}\label{eq_calhN}
\end{equation}
describes the charge carrier with momentum $k$, spin $\sigma$ in the normal metal and
\begin{equation}
\calh_{S}=\sum_{k\sigma}\varepsilon_{Sk\sigma}c_{Sk\sigma}^{\dag}c_{Sk\sigma}+\sum_{k}\left[\Delta c_{S,-k\up}^{\dag}c_{Sk\down}^{\dag}+{\rm H.c.}\right]
\end{equation}
describes the superconductor with an order parameter given by the energy gap $\Delta$. The temperature dependence of the gap can be approximated by \cite{kamp_phase_2019}
\beq\label{eq_DT}
\Delta(T)=\Delta_0\tanh\left(1.74\sqrt{\frac{T_c}{T}-1}\right)
\edq
with $k_BT_c=0.568\Delta_0$ with the critical temperature $T_c$. The approximation given by Eq.~\eqref{eq_DT} deviates by less than two percent from the solution of the self-consistency equation for the order parameter. Since there is only one superconductor in our system, we can choose $\Delta$ to be real.

We model the quantum dot as a spin-degenerate energy level $\varepsilon_{d}$, which can be controlled by a gate voltage.
Further, we treat the Coulomb interaction at the quantum dot in the mean-field approximation, which facilitates the incorporation of nonlinear effects self-consistently \cite{christen1996gauge}.
This is justified since the cooling effects in this work mainly stem from detuning the quantum dot energy level rather than from the interaction. However, screening potential enhances the refrigeration power, as we show below. Therefore, the dot Hamiltonian in Eq.~\eqref{eq_calh} is given by
\begin{equation}\label{HD}
\calh_{D}=\sum_{\sigma}(\varepsilon_{d}+U)d_{\sigma}^{\dag}d_{\sigma}\,,
\end{equation}
where the mean-field Coulomb interaction strength $U(V)$ can be a nonlinear function of voltage to an arbitrary order \cite{sanchez_scattering_2013,meair2013scattering,lopez2013nonlinear}.
For small quantum dots, which is the experimentally relevant situation, screening effects are strong and $U(V)$ can be found by solving \cite{hwang_hybrid_2016}
\beq\label{rho}
\delta \rho(V,U)=0\,
\edq
where $\delta \rho=\rho-\rho_{\text{eq}}$ is the charge density fluctuation.

Finally, the charge tunneling between the quantum dot and each lead is described  in Eq.~\eqref{eq_calh}  by
\begin{equation}
\calh_{T}=\sum_{k\sigma}t_{N\sigma}c_{Nk\sigma}^{\dag}d_{\sigma}+\sum_{k\sigma}t_{S\sigma}c_{Sk\sigma}^{\dag}d_{\sigma}+{\rm H.c.}\,,
\end{equation}
where $t$ are the tunnel amplitudes between the different subsystems.

We evaluate the charge and heat currents \cite{hwang_hybrid_2016} from the time evolution of  electron number $N=\sum_{k\sigma}c_{Nk\sigma}^{\dag}c_{Nk\sigma}$ and the energy in the left metal given by Eq. \eqref{eq_calhN},
\begin{align}
&I=-(ie/\hbar)\bra[\calh,N]\ket\,,\label{eq:charge}\\
\label{eq:Jsigma}&J=-(i/\hbar)\bra[\calh,\calh_{N}]\ket-I V\,,
\end{align}
where the last term in Eq.~\eqref{eq:Jsigma} corresponds to the Joule heating dissipated by Andreev and quasiparticle currents, $I=I_A+I_Q$, as derived below.

We apply the nonequilibrium Keldysh-Green formalism to the heat current~\cite{hwang_large_2016}, where we find two separate contributions $J=J_{A}+J_{Q}$ with
\begin{align}
&J_{A}=-2VI_{A}\label{J_A}\,,\\
&J_{Q}=\frac{2}{h}\int d\varepsilon~(\varepsilon-eV)~T_{Q}(\varepsilon)\big[f_{L}(\varepsilon-eV)-f_{R}(\varepsilon)\big]\label{J_Q}\,,
\end{align}
where $f_{L}(\varepsilon\pm eV)=\{1+\exp[(\varepsilon\pm eV-E_F)/k_{B}T_{N}]\}^{-1}$ is the Fermi-Dirac distribution at the left metal with a bias voltage $V$ and temperature $T_N$, while $f_R(\varepsilon)=\{1+\exp[(\varepsilon-	E_F)/k_{B}T]\}^{-1}$ corresponds to that in the right superconductor
($T$ is the background temperature).
Equation~\eqref{J_A} is the Joule heating from the Andreev current $I_A$ which is detrimental to the refrigeration effect dominantly at subgap transport. This Joule heating as well as that of quasiparticle contribution appear only in the nonlinear regime. This shows the importance of a fully nonlinear theory of transport as employed here. Note that the factor 2 in Eq. \eqref{J_A} comes from the particle and hole contributions while that in Eq. \eqref{J_Q} for quasiparticles reflects the spin degeneracy. Equation \eqref{J_Q} can be decomposed into $J_Q=J_Q^E-I_QV$, where
\beq\label{JQE}
J_{Q}^E=\frac{2}{h}\int d\varepsilon~\varepsilon T_{Q}(\varepsilon)\big[f_{L}(\varepsilon-eV)-f_{R}(\varepsilon)\big]\,,
\edq
is the energy current carried by quasiparticles and
\beq
I_{Q}=\frac{2e}{h}\int d\varepsilon~T_{Q}(\varepsilon)\big[f_{L}(\varepsilon-eV)-f_{R}(\varepsilon)\big]\label{I_Q}
\edq
is the quasiparticle electric current. Notably, the Andreev heat current in Eq. \eqref{J_A} does not contain the energy current unlike $J_Q$. This is due to the intrinsic particle-hole symmetry present in the subgap regime. Indeed, the Andreev spectrum, cf. Eq. \eqref{eq:TA}, is always particle-hole symmetric, i.e., $T_A(\varepsilon)=T_A(-\varepsilon)$, even if the finite gate voltage is applied to the quantum dot. This is in stark contrast to the quasiparticle transmission where $T_Q(\varepsilon)\ne T_Q(-\varepsilon)$ with $\varepsilon_d\ne0$. The latter fact is crucial for our refrigerator to function properly.

The Andreev charge current in Eq. \eqref{J_A} is explicitly written by (here the factor 2 is due to spin)
\beq
I_{A}=\frac{2e}{h}\int d\varepsilon~T_{A}(\varepsilon)\big[f_{L}(\varepsilon-eV)-f_{L}(\varepsilon+eV)\big]\label{I_A}\,.
\edq
One can write the Andreev and quasiparticle transmission functions in the above equations in terms of the retarded Green's functions 
\begin{align}
&T_{A}(\varepsilon)=\Gamma_{L}^2\big|G_{12}^{r}\big|^{2}\,,\label{eq:TA}\\
&T_{Q}(\varepsilon)=\Gamma_{L}\widetilde{\Gamma}_R\Big(\big|G_{11}^{r}\big|^{2}+\big|G_{12}^{r}\big|^{2}-\frac{2\Delta}{|\varepsilon|}\text{Re}\big[G_{11}^rG_{12}^{r,*}\big]\Big)\,,\label{eq:TQ}
\end{align}
where $\widetilde{\Gamma}_R=\Gamma_{R}\Theta(|\varepsilon|-\Delta)|\varepsilon|/\sqrt{\varepsilon^{2}-\Delta^{2}}$, and
\begin{align}
&G^{r}_{11}=\Big[\varepsilon-\varepsilon_{d}+\frac{i\Gamma_{L}}{2}+\frac{i\Gamma_{R}}{2}\beta_d(\varepsilon)+\frac{\Gamma_R^{2}\Delta^{2}}{4(\varepsilon^2-\Delta^{2})}A_1^r(\varepsilon)\Big]^{-1},\\
&G^{r}_{12}=G^{r}_{11}\frac{i\Gamma_{R}}{2}\beta_o(\varepsilon)A_1^r(\varepsilon)\,,
\end{align}
with
\begin{align}
&A_1^r(\varepsilon)=\Big[\varepsilon+\varepsilon_{d}+\frac{i\Gamma_{L}}{2}+\frac{i\Gamma_{R}}{2}\beta_d(\varepsilon)\Big]^{-1},\\
&\beta_d(\varepsilon)=\frac{\Theta(|\varepsilon|-\Delta)|\varepsilon|}{\sqrt{\varepsilon^{2}-\Delta^{2}}}-i\frac{\Theta(\Delta-|\varepsilon|)\varepsilon}{\sqrt{\Delta^{2}-\varepsilon^{2}}}\,,\\
&\beta_o(\varepsilon)=\frac{\Theta(|\varepsilon|-\Delta)\text{sgn}(\varepsilon)\Delta}{\sqrt{\varepsilon^{2}-\Delta^{2}}}-i\frac{\Theta(\Delta-|\varepsilon|)\Delta}{\sqrt{\Delta^{2}-\varepsilon^{2}}}\,.
\end{align}

The charge density fluctuation in Eq. \eqref{rho} can be evaluated from the lesser Green's function
\beq
\delta\rho=-i\int d\varepsilon\Big[G_{11}^{<}(\varepsilon)-G_{11,\text{eq}}^{<}(\varepsilon)\Big]\,,
\edq
with
\beq\label{Glesser}
\begin{split}
G_{11}^{<}&(\varepsilon)=\frac{i\Gamma_{L}}{2\pi}\bigg[\big|G_{11}^{r}\big|^{2}f_{L}(\varepsilon-eV)
	+\big|G_{12}^{r}\big|^{2}f_{L}(\varepsilon+eV)\bigg]\\
&+\frac{i\widetilde{\Gamma}_{R}}{2\pi}f_{R}(\varepsilon)\bigg[\big|G_{11}^{r}\big|^{2}
	+\big|G_{12}^{r}\big|^{2}-\frac{2\Delta}{|\varepsilon|}\text{Re}\big[G_{11}^rG_{12}^{r,*}\big]\bigg]\,,
\end{split}
\edq
where $G_{11,\text{eq}}^{<}(\varepsilon)$ is the value of $G_{11}^{<}(\varepsilon)$ taken at zero voltage and temperature biases.
In the derivation of Green's functions, we have used the wide band approximation, i.e., energy-independent tunnel couplings. As a consequence,
the coupling strengths read $\Gamma_{L/R}=2\pi\sum_{k\sigma} |t_{N/S\sigma}|^2 \delta (\varepsilon-\varepsilon_{N/Sk\sigma})$.

Once the total heat flow $J=J_Q^E-I_QV-2I_AV$ removes or adds the heat to the normal metal, it subsequently decreases or increases the temperature for the normal metal $T_N$, which we take initially as $T_N=T$. Thus,
\beq\label{DT}
T_N=T+\delta T\,,
\edq
where $\delta T$ is the change of temperature at the left normal metal side due to heat removal or addition. In order to calculate the temperature change $\delta T$, we consider the energy flow between the normal reservoir and the phonons in the substrate via the phenomenological equation based on the simple thermal model \cite{wellstood_hot_1994}
\beq\label{eq:el-ph}
P_{\text{el-ph}}=\Sigma{\mathcal{V}}(T_N^5-T_\text{ph}^5)\,,
\edq
where $\Sigma$ represents the electron-phonon coupling strength and ${\mathcal{V}}$ denote the volume of the conductor. Here, $T_\text{ph}$ denotes the temperature of the phonon which we take the same as substrate temperature in the calculation. For typical metals, one can assume $\Sigma\simeq10^9$~WK$^{-5}$m$^{-3}$ \cite{giazotto_opportunities_2006}. We also take ${\mathcal{V}}\simeq10^{-20}$m$^{3}$, which is a reasonable value for generic mesoscopic devices. $\delta T$ can then be evaluated from the heat balance equation
\beq\label{balance}
J+P_{\text{el-ph}}=0\,.
\edq
Now the temperature change $\delta T$ generates a nonequilibrium thermal bias between superconducting and normal leads and hence an emerging thermocurrent.
In turn, $J$ changes according to Eq. \eqref{eq:Jsigma}. Additionally, the temperature shift at the normal metal affects the charge density distribution via Eq. \eqref{Glesser} and the solution to Eq. \eqref{rho} is also modified, which leads to a further change of $T_N$ via Eqs. \eqref{DT}, \eqref{eq:el-ph}, and \eqref{balance}. This iteration process continues until the stationary heat flow and the temperature are reached. 
In what follows, we will discuss the heat flow and temperature of the cooled reservoir at the steady state for which the convergence of our self-consistent calculation
is achieved. 

\begin{figure}[t]
\centering
\includegraphics[width=0.35\textwidth]{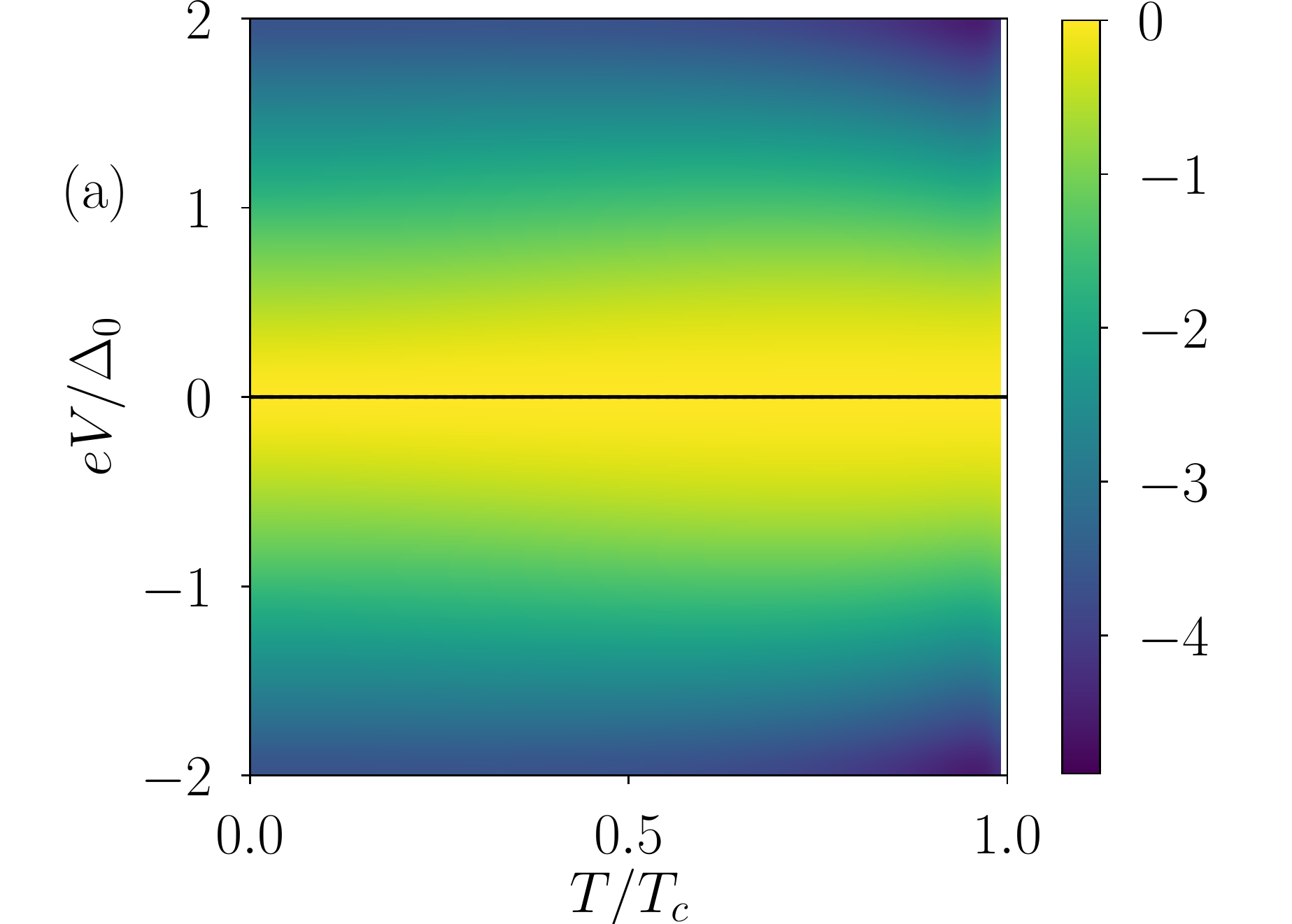}\\
\includegraphics[width=0.35\textwidth]{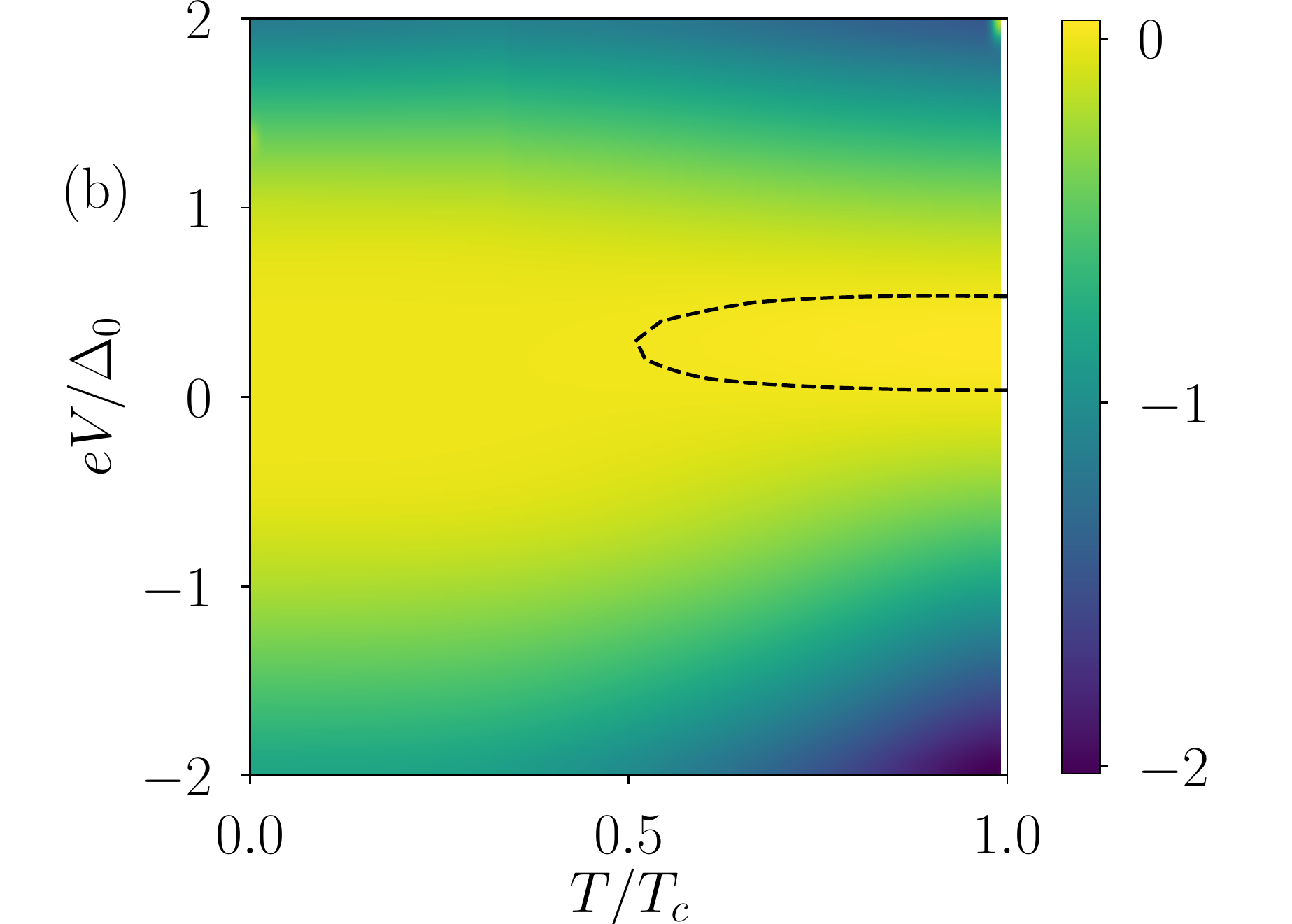}\\
\includegraphics[width=0.35\textwidth]{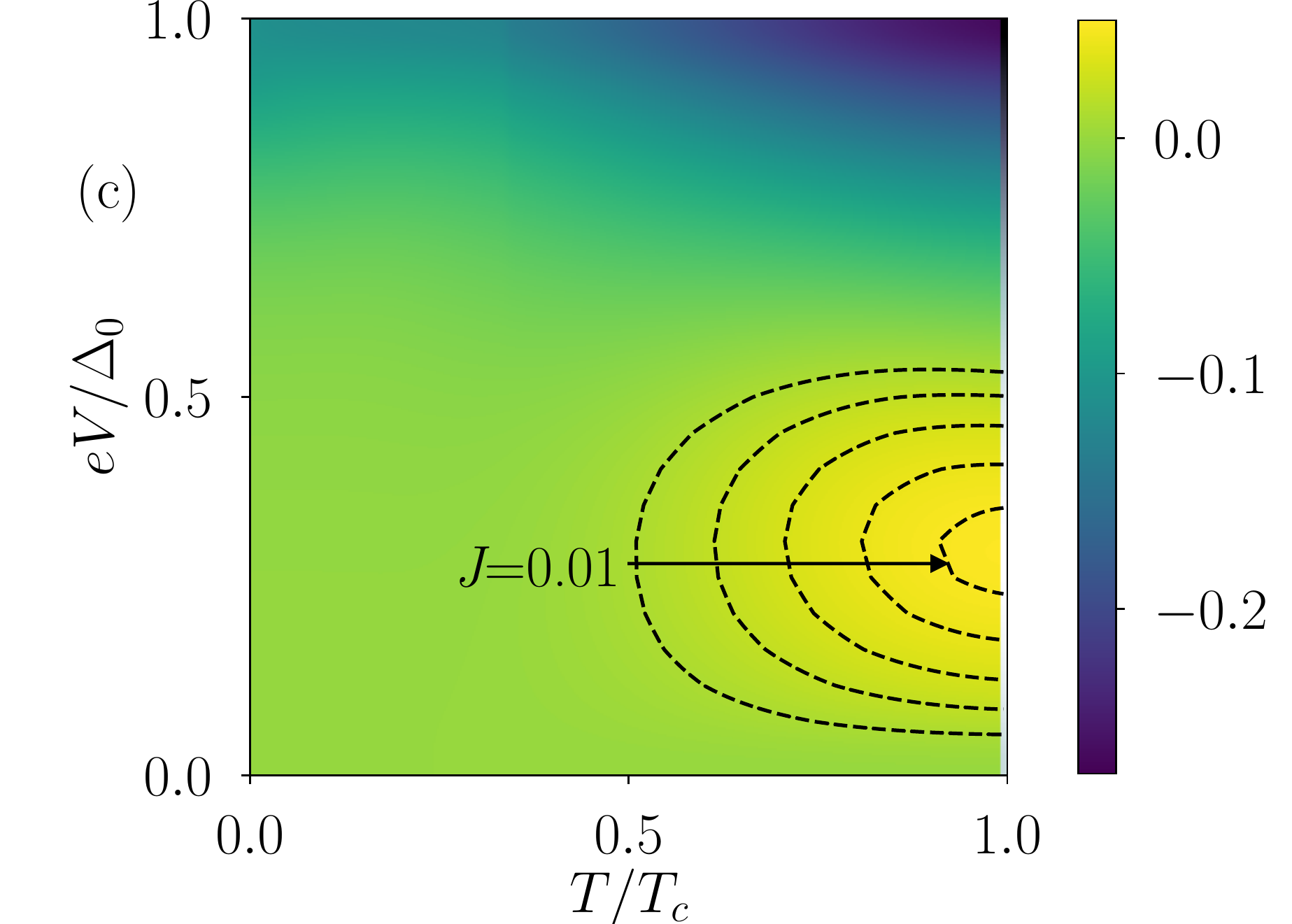}
  \caption{Density plot of $J/[\Delta_0^2/h]$ with (a) $\veps_d=0$, (b) $\veps_d=1.5\Delta_0$. (c) Detailed map of (b) in the positive voltage region $0<eV<\Delta_0$ where the refrigeration regime can be identified. No cooling appears in (a), i.e., $J<0$. In (a), the black line denotes $J(V=0)=0$ while in (b), $J=0.01$ contour line is drawn. In (c), the arrow represents the increasing cooling power in steps of 0.01 starting from $J=0.01$. Coupling parameters are $\Gamma_L=\Gamma_R=0.5\Delta_0$. $T_c$ is the critical temperature, and $T$ is the temperature of superconductor which is assumed to agree with the background temperature, as explained in the main text. }
  \label{Fig2}
\end{figure}
\begin{figure}[htbp]
\centering
\includegraphics[width=0.4\textwidth]{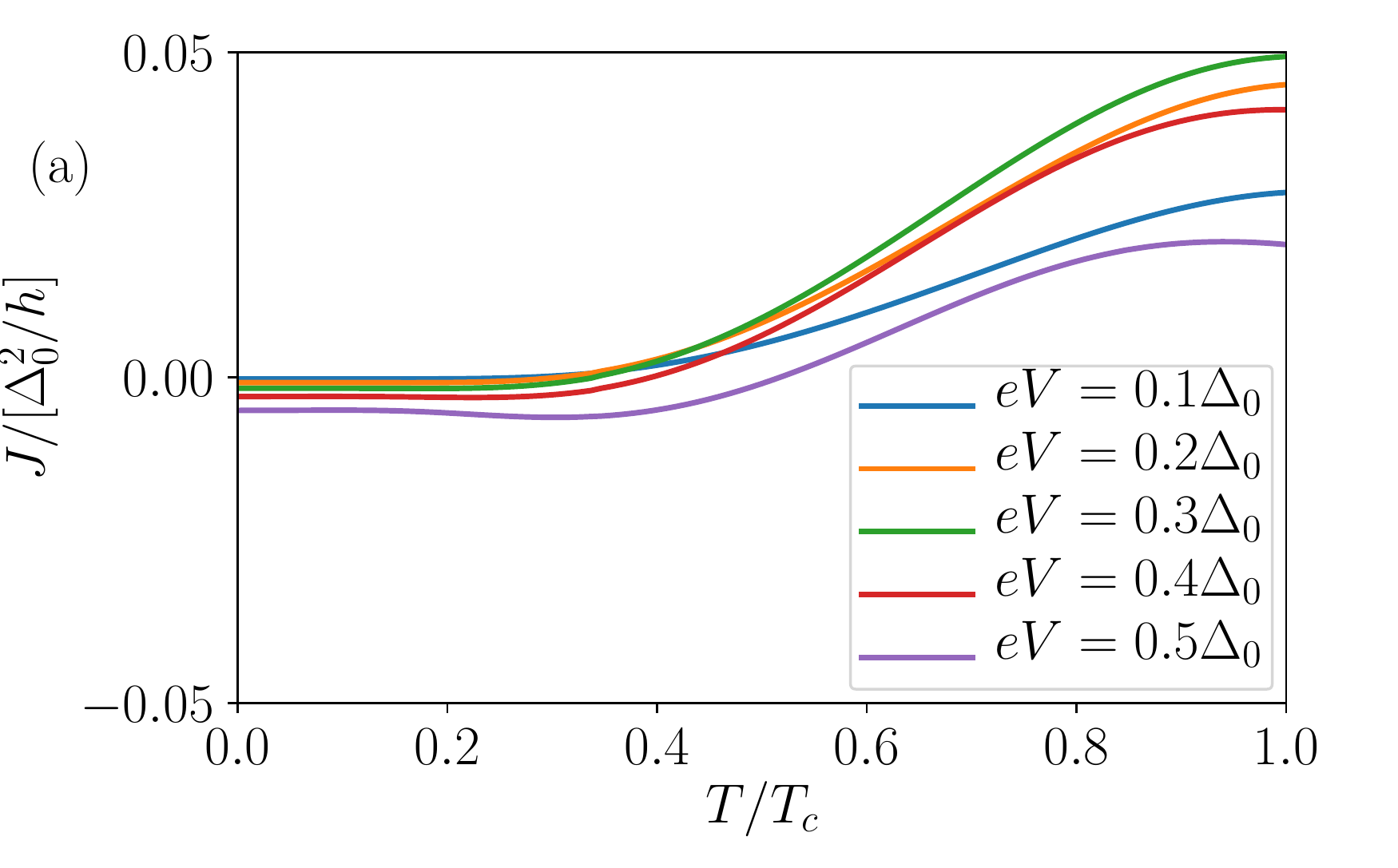}\\
\includegraphics[width=0.4\textwidth]{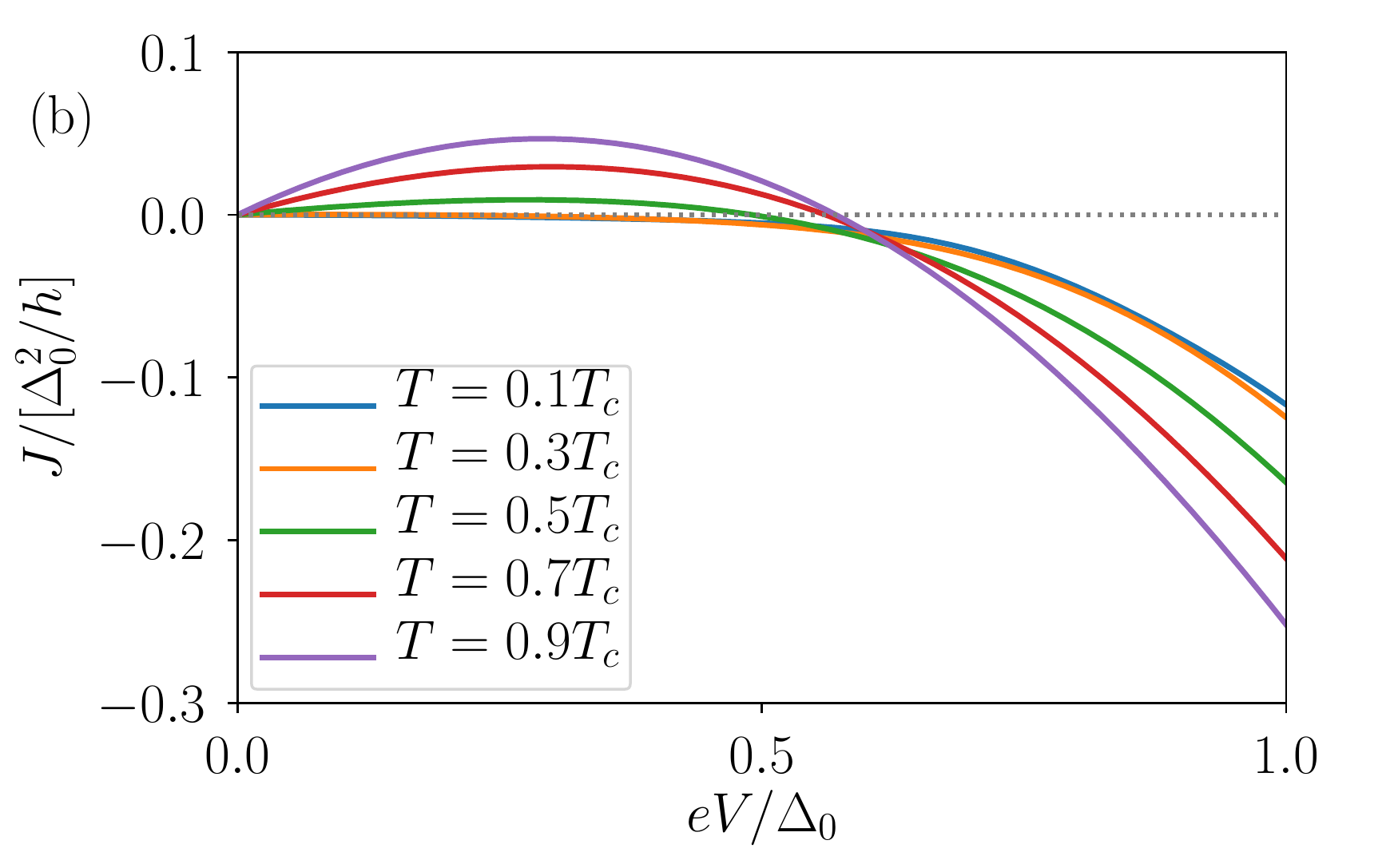}
  \caption{$J$ versus (a) $T$ for several $V$ and (b) $eV$ for several $T$ at $\veps_d=1.5\Delta_0$ [cf. Fig. \ref{Fig2}(c)]. Coupling parameters are $\Gamma_L=\Gamma_R=0.5\Delta_0$. In (b), negative voltage side only shows strong heating, i.e., $J(V<0)<0$.}
  \label{Fig3}
\end{figure}
\begin{figure*}[htbp]
  \begin{tabular}{cc}
    \includegraphics[width=0.35\textwidth]{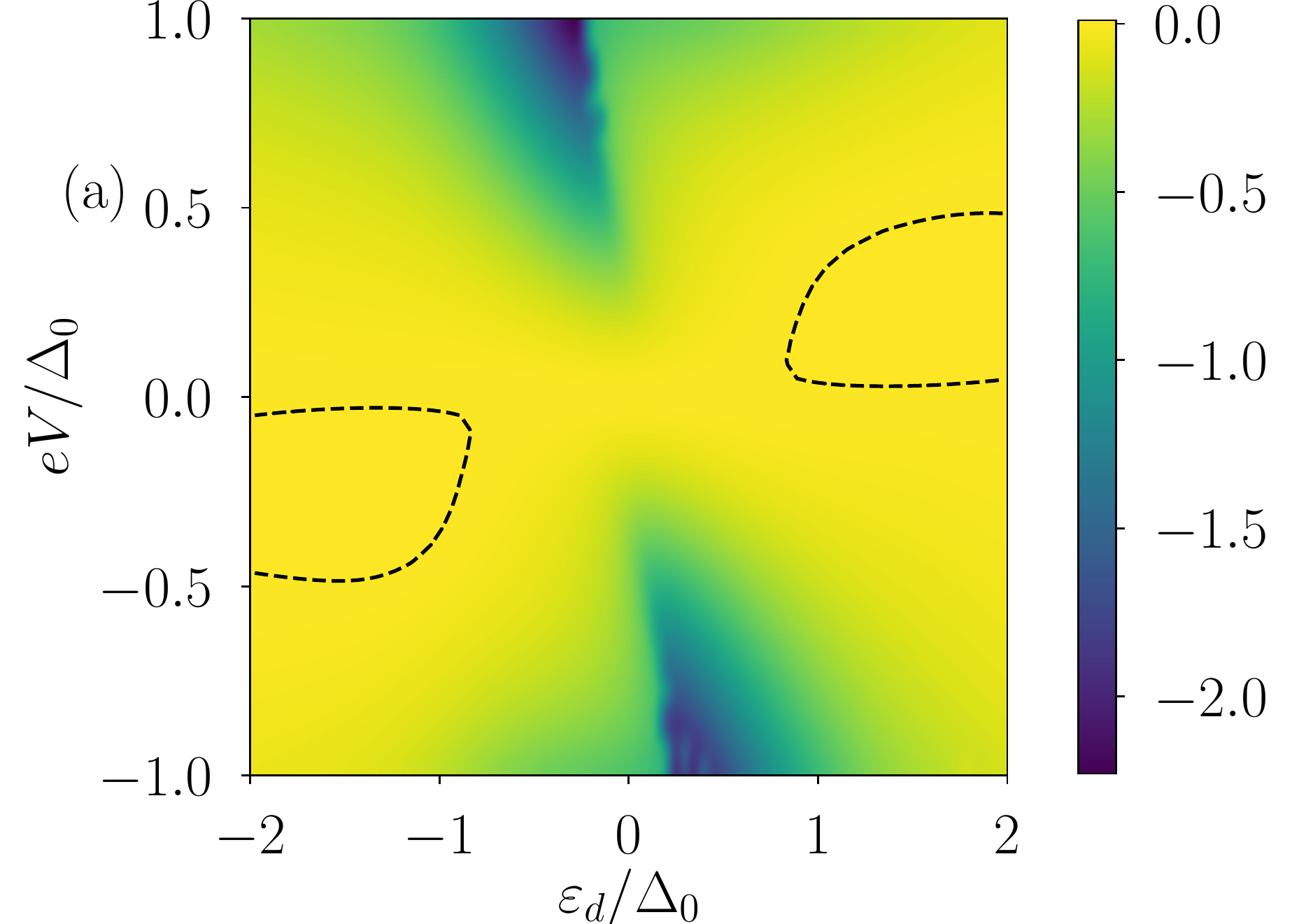}&
\includegraphics[width=0.35\textwidth]{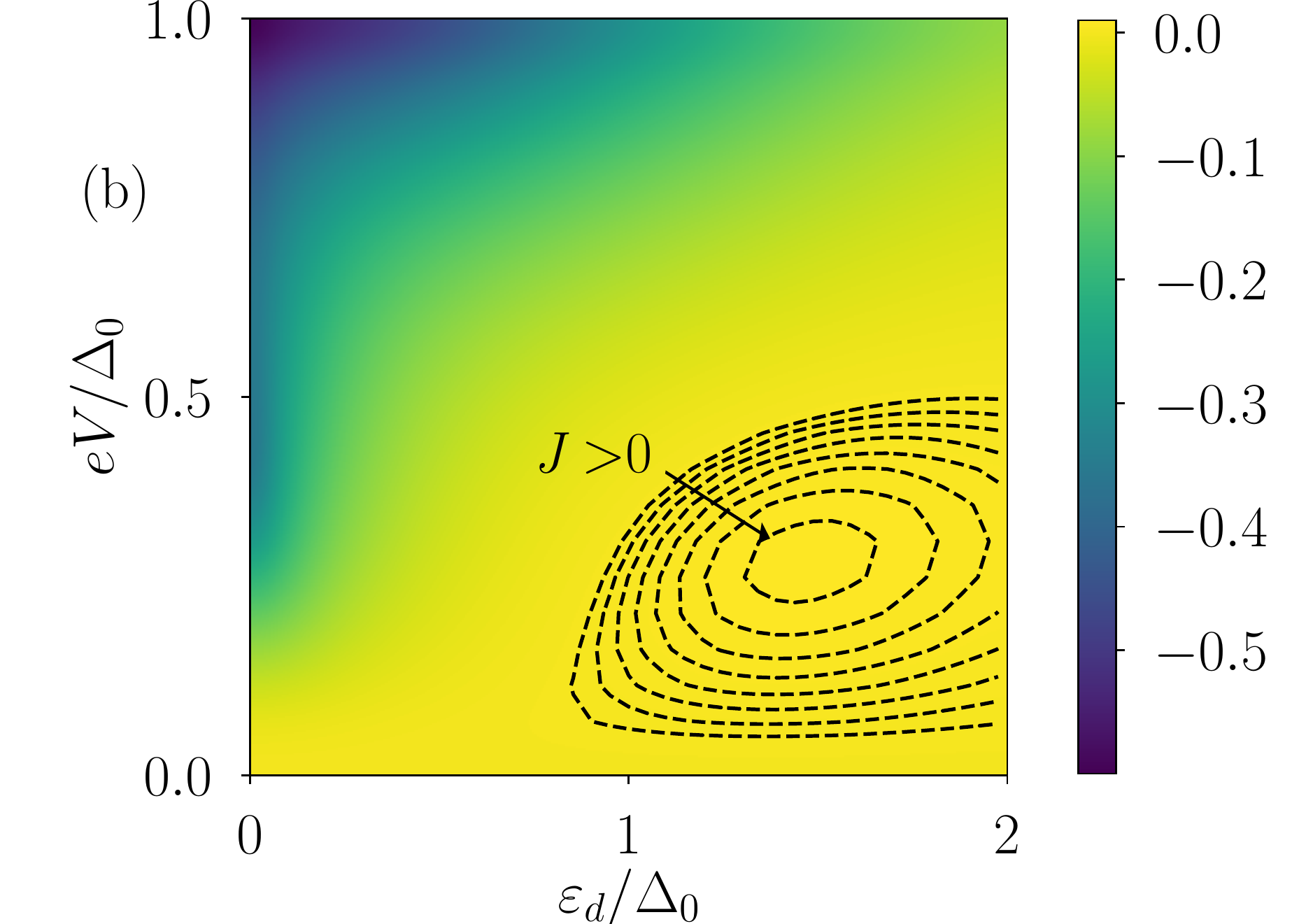}\\
\includegraphics[width=0.35\textwidth]{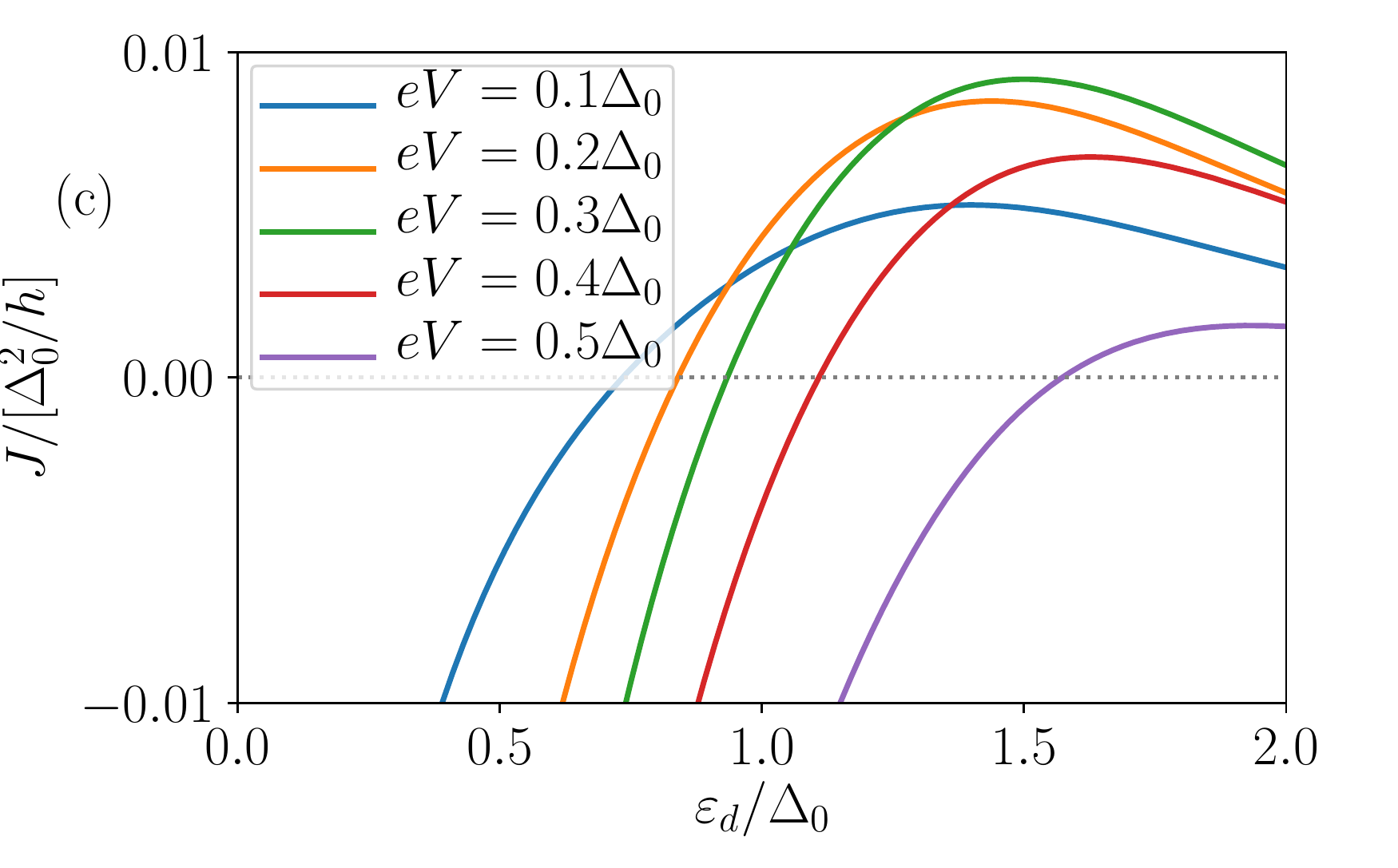}&
\includegraphics[width=0.35\textwidth]{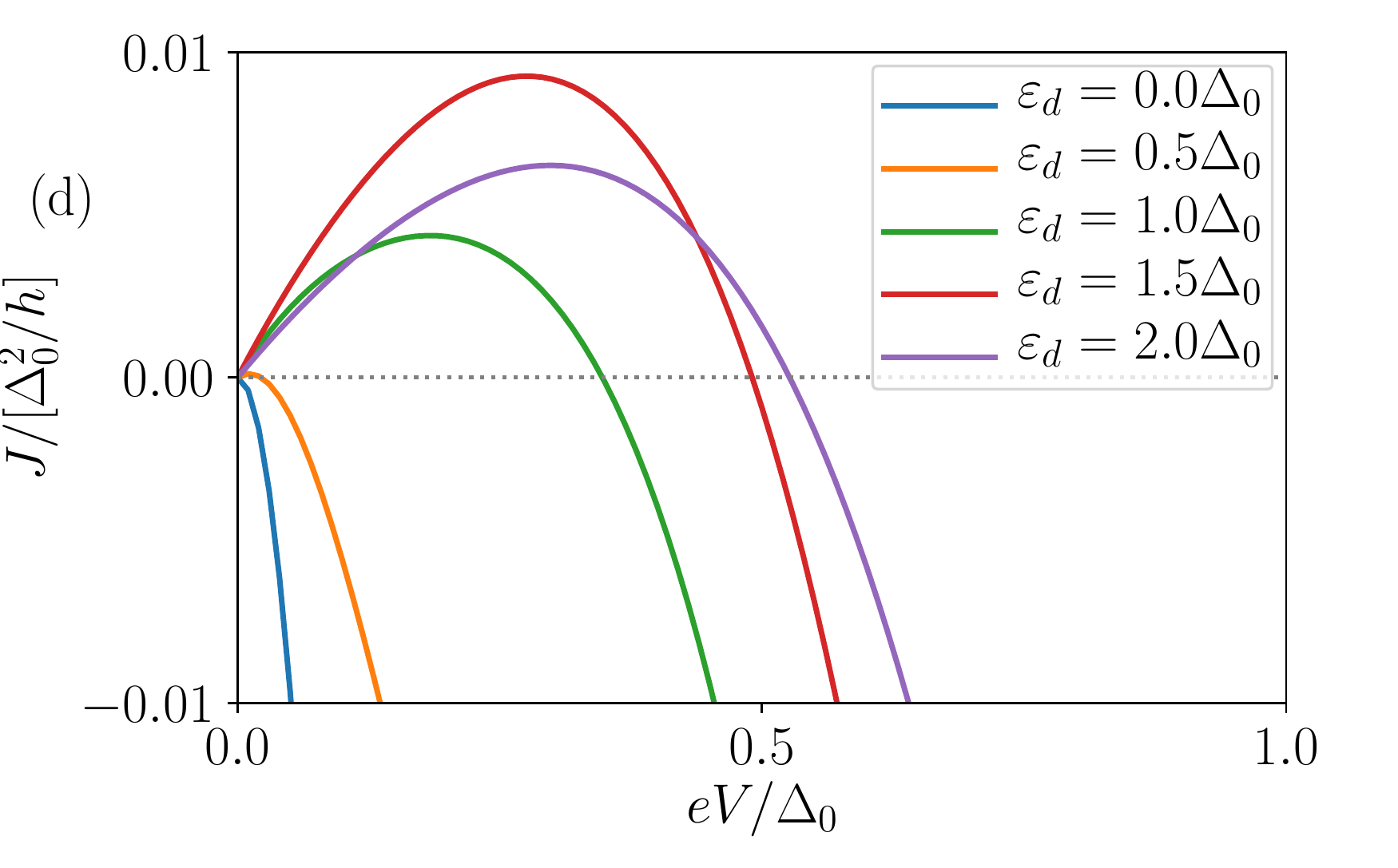}
  \end{tabular}
  \caption{(a) Density plot of $J/[\Delta_0^2/h]$ at $T=0.5T_c$ with $J=0.002$ contour drawn with the dashed line, (b) detailed map of (a) in the positive gate and voltage region where the higher refrigeration regime can be identified with the contour lines from $J=0.002$ to $J=0.009$ in steps of 0.001. (c) $J$ vs $\veps_d$, (d) $J$ vs $eV$. Minimum condition for cooling is $V\veps_d>0$. Maximum cooling of $J\sim0.01\Delta_0^2/h$ appears around $\veps_d=1.5\Delta_0$ and $eV=0.3\Delta_0$. Coupling parameters are $\Gamma_L=\Gamma_R=0.5\Delta_0$.}
  \label{Fig4}
\end{figure*}

\section{Results and discussion}\label{results}
Figure  \ref{Fig2} shows the density map of the stationary heat flow $J$ as a function of applied voltage ($V$) and background temperature ($T$).
In Fig. \ref{Fig2}(a), we observe that no cooling effect occurs, i.e, $J<0$, in the whole temperature range of consideration $0<T<T_c$ for the particle-hole symmetry point $\veps_d=0$. In this case, Joule heating is dominant and hence the applied voltage bias, both in positive ($V>0$) and negative ($V<0$) directions, only heats the normal metal reservoir. In order to achieve the refrigeration via Peltier effect, i.e., by applying the voltage bias to the normal lead, it is neccessary to break the particle-hole symmetry. In our system, detuning the quantum dot level by applying a gate voltage can easily break this symmetry in the quasiparticle spectrum in Eq. \eqref{eq:TQ}. Applied voltage at the normal metal induces screening potential $U$ which additionally contributes to the dot level shift. At optimum conditions, these effects can lead to large positive values of the quasiparticle cooling, i.e., $J_Q>0$ in Eq. \eqref{J_Q}. However, this effort to find large cooling effects should overcome major obstacles. First, in Eq. \eqref{J_Q}, the energy current $J_Q^E$, viz. Eq. \eqref{JQE}, should be larger than the Joule heating counterpart $I_QV$ driven by quasiparticles so that the net heat flow $J_Q^E-I_QV$ can be positive. Next, the Joule heating from Andreev current, viz. Eq. \eqref{J_A}, should not surpass the cooling effect from the quasiparticles. Finally, once the voltage is applied to the normal metal from the isothermal situation, i.e., $T_N=T$, heat is removed or added due to Peltier effect. If its thermocurrent contribution is initially large, one ends up with net heating effect ($J<0$) with an elevated temperature $T_N>T$, i.e., $\delta T>0$ in Eq. \eqref{DT}, when the steady state is reached. In this case, the quasiparticle refrigeration effect is hindered by this thermoelectric effect. At optimum conditions, however, the induced thermocurrent can be minimal and hence one can finally achieve net cooling with the stationary heat flow $J>0$ with a temperature of the normal metal lower than the original value, i.e., $T_N<T$.

In Fig. \ref{Fig2}(b), when $\veps_d\ne0$ as exemplified with $\veps_d=1.5\Delta_0$, cooling can indeed become dominant, i.e., $J>0$, for a certain range of positively applied voltage bias and background temperature while the heating effect dominates outside this optimum range. The sign of this optimum voltage range and the dot level detuning satisfy the condition $V\veps_d>0$ (we will later explain this effect).
Figure \ref{Fig2}(c) shows the positive voltage range $0<eV<\Delta_0$ of Fig. \ref{Fig2}(b) where the maximum cooling of $J\sim 0.05\Delta_0^2/h$ appears with $eV\sim 0.3\Delta_0$.
This rather large cooling power can be realized by tuning the quantum dot level via gate voltages. In Ref. \cite{rouco_electron_2018}, a cooling power of the order of $0.05\Delta_0^2/h$ is reported in the presence of a Zeeman field. The devices in Ref.~\cite{rouco_electron_2018} exhibit a refrigeration effect already in linear response in combination with spin filtering barriers. Our setup operates in the nonlinear regime but, more importantly, works in fully electric ways without any magnetic components. Fig. \ref{Fig2}(c) zooms in Fig. \ref{Fig2}(b) to show that refrigeration effects are absent at low voltages and temperatures. In this case, the quasiparticle cooling itself is rather small, i.e., both thermal excitation of quasiparticles and the Peltier effect are suppressed. Instead, the Andreev Joule heating $J_A$ is larger than $J_Q$ leading to net heating  $J_Q^E-I_QV-2I_AV<0$. In contrast, at higher voltages and temperatures, $J_A$ can be reduced but the Joule heating from the quasiparticles $I_QV$ becomes larger than its energy current counterpart $J_Q^E$ hence diminishing the net refrigeration, i.e., $J_Q^E-I_QV-2I_AV\approx J_Q^E-I_QV<0$.

Let us make a closer inspection of the refrigeration effect found in Fig.~\ref{Fig2}(c).
Figure \ref{Fig3}(a) displays $J$ versus $T$ for selected values of $eV$ with $\Gamma_L=\Gamma_R=0.5\Delta_0$. The cooling effect is suppressed for $T<0.4T_c$ as the voltage is applied. As explained above, this is caused by  the quasiparticle energy current $J_Q^E$ inability to overcome the net Joule heating generated by the Andreev and quasiparticle charge currents. At higher temperatures $T>0.5T_c$, the cooling power is comparatively small for low voltages, e.g., $eV=0.1\Delta_0$. However, as higher voltage is applied, one can increase the refrigeration effects up to a certain point, e.g., $eV=0.3\Delta_0$. Then, the cooling power becomes smaller again for higher voltages as shown with $eV=0.5\Delta_0$. Indeed, one can notice that there is an optimum voltage above which the cooling power starts to decrease. This is more easily visible in Fig. \ref{Fig3}(b) where we plot $J$ as a function of $eV$ for several $T$. Remarkably, the refrigeration effect persists in a rather broad range of voltage as temperature approaches $T_c$. 
For $eV>0.6\Delta_0$, as seen in Fig. \ref{Fig3}(b), one clearly notices that there are only net heating effects irrespective of the base temperatures. Moreover, for higher temperatures, e.g., $T=0.7T_c$ and $T=0.9T_c$, both cooling power at lower voltages and heating effects at higher voltages are strong indicating the role of quasiparticles in both cases. For the former, $J_Q^E$ plays a significant role for refrigeration with an optimum voltage bias range, whereas $I_QV$ becomes dominant for the latter with higher voltage. At a slightly lower temperature $T=0.5T_c$, both cooling and heating powers are smaller in magnitude in comparison to those at higher temperatures. At very low temperatures, e.g., $T=0.1T_c$ and $T=0.3T_c$, no refrigeration effect is observed as $J_Q^E$ becomes strongly suppressed. In this case, quasiparticles are mostly inactive and hence the heating powers are also smaller than those at elevated temperatures.
It should be mentioned that while the cooling power $J$ becomes larger as $T$ approaches close to $T_c$ [Fig. \ref{Fig3}(a)], the corresponding temperature drop $\delta T$ due to Peltier effect is the largest well below $T_c$ around $T\approx 0.6T_c$, cf. Fig. \ref{Fig6}, indicating the importance of superconductivity with the strongly energy-dependent density of states.

\begin{figure*}[htbp]
  \begin{tabular}{cc}
    \includegraphics[width=0.35\textwidth]{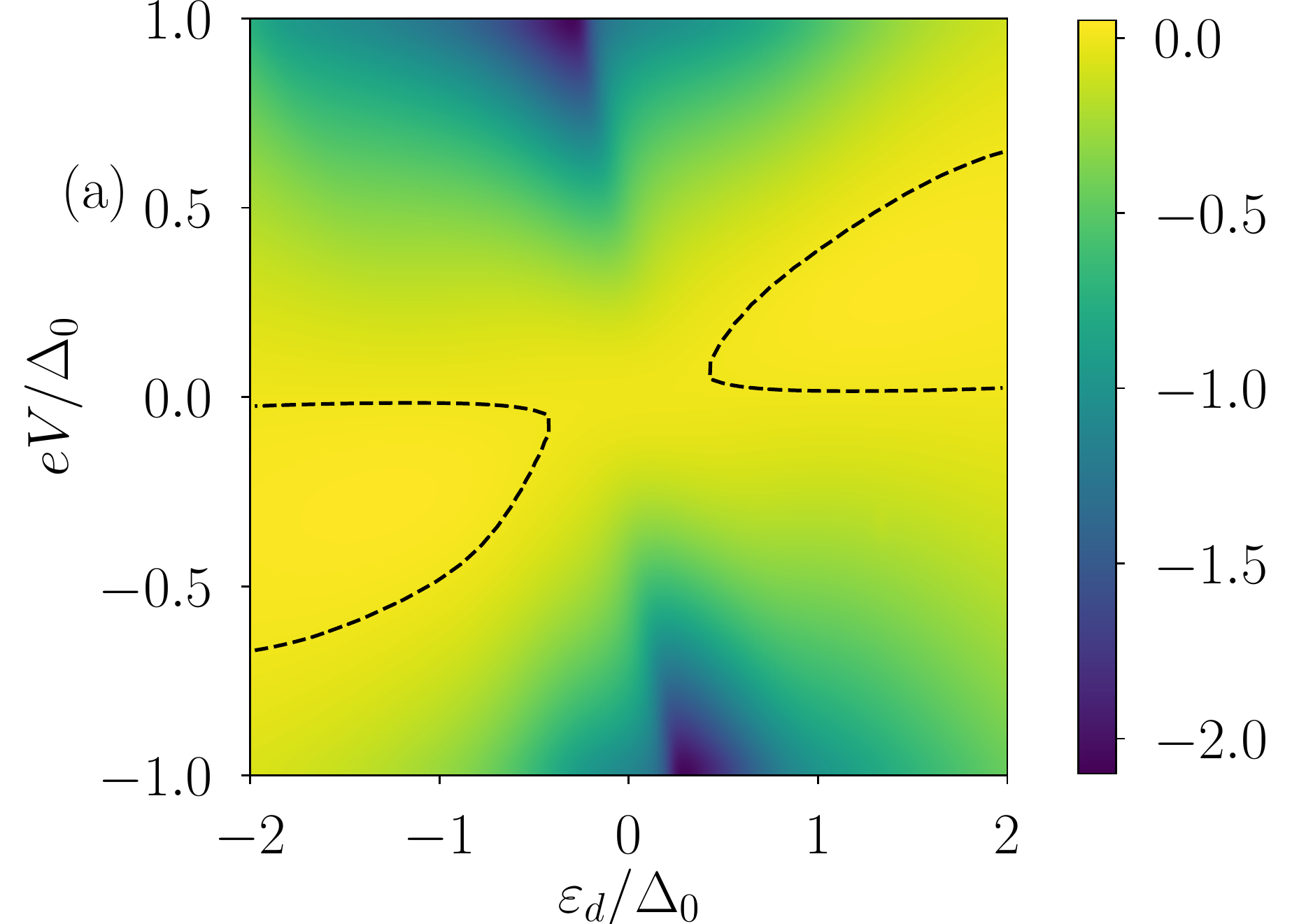}&
\includegraphics[width=0.35\textwidth]{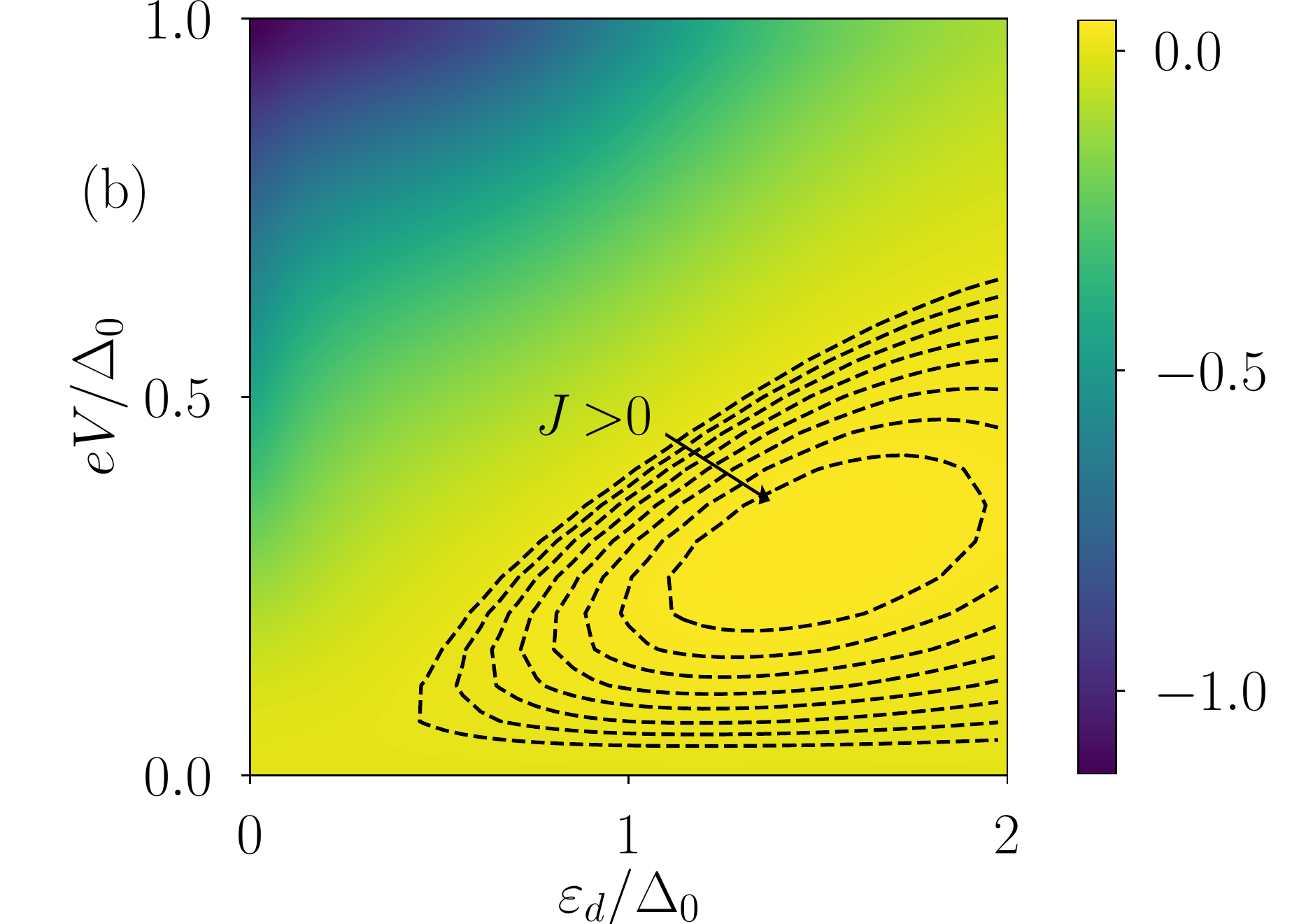}\\
\includegraphics[width=0.35\textwidth]{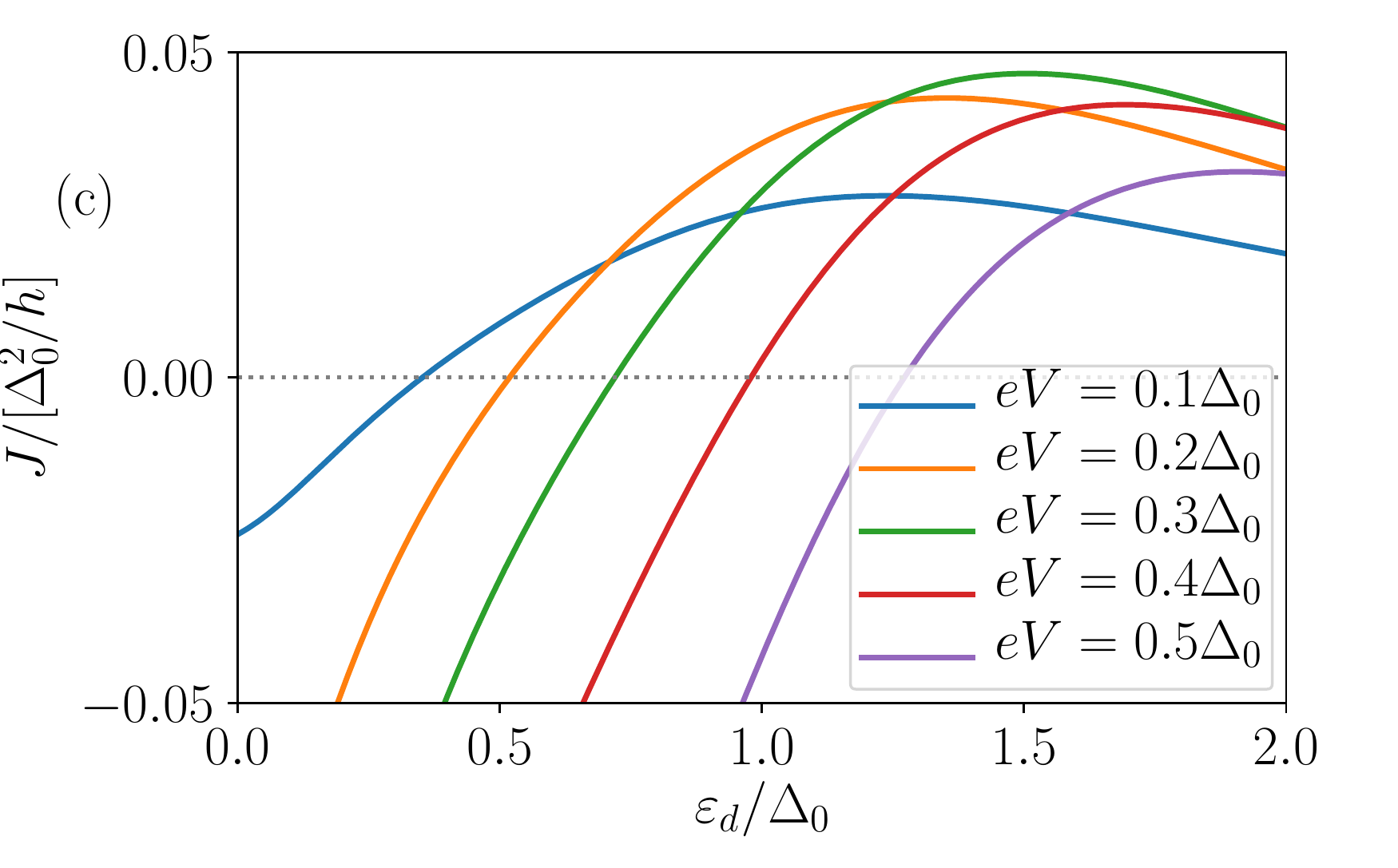}&
\includegraphics[width=0.35\textwidth]{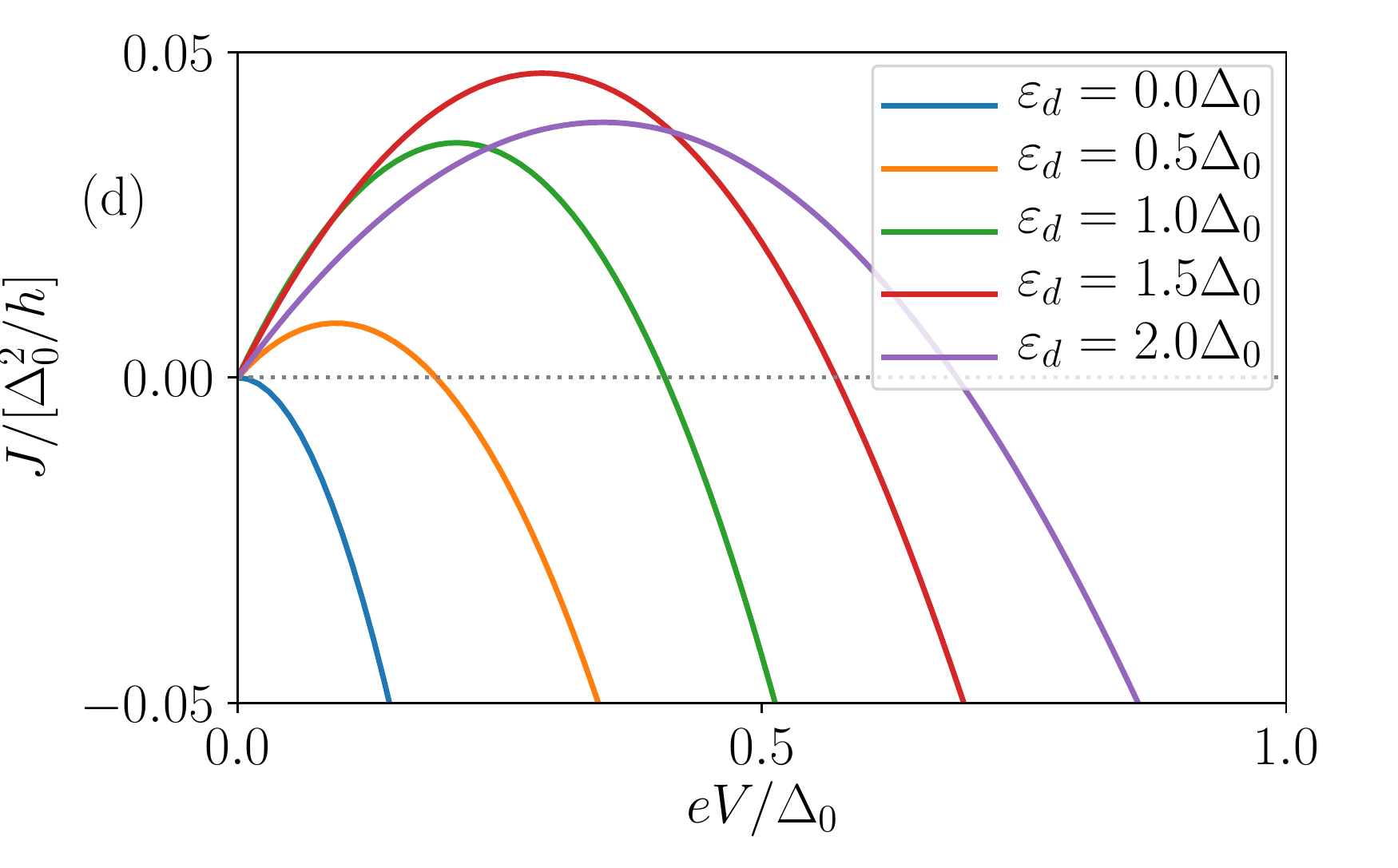}
  \end{tabular}
  \caption{(a) Density plot of $J/[\Delta_0^2/h]$ at $T=0.9T_c$ with $J=0.005$ contour drawn with the dashed line, (b) detailed map of (a) in the positive gate and voltage region where the higher refrigeration regime can be identified with the contour lines from $J=0.005$ in steps of 0.005. (c) $J$ vs $\veps_d$, (d) $J$ vs $eV$. Minimum condition for cooling is $V\cdot\veps_d>0$. Maximum cooling of $J\sim0.05\Delta_0^2/h$ appears around $\veps_d=1.5\Delta_0$ and $eV=0.3\Delta_0$. Coupling parameters are $\Gamma_L=\Gamma_R=0.5\Delta_0$.}
  \label{Fig5}
\end{figure*}

We have thus far illustrated the refrigeration effect at a fixed gate potential $\veps_d=1.5\Delta_0$. We now discuss the effect of varying $\veps_d$ fixing the background temperature. Figure \ref{Fig4}(a) illustrates a density plot of $J$ as a function of both the quantum dot level position $\veps_d$ and the applied voltage $V$. The temperature is set to be $T=0.5T_c$.
The dashed lines indicate the optimum parameter range within which the net cooling effects can appear. As shown in Fig. \ref{Fig4}(a) [see also Fig. \ref{Fig5}(a) below], we find that there are two refrigeration regimes once the quantum dot energy level and the voltage bias have the same signs, as earlier anticipated. Indeed, we have an approximate symmetry such that $J(\veps_d,V)\approx J(-\veps_d,-V)$. However, it should be emphasized that this is not an exact symmetry since the Hartree approximation of the screening potential given by Eq. \eqref{rho} breaks the particle-hole symmetry. The reason for this approximate symmetry is that the screening potential in our model satisfies $U(\veps_d,V)\approx-U(-\veps_d,-V)$ \cite{lim2010magnetoasymmetric}. Thus, upon simultaneous reversal of $\veps_d$ and $V$, the renormalized quantum dot level $\veps_d+U$ in Eq. \eqref{HD} becomes $-(\veps_d+U)$. The numerical values obtained by solving the self-consistent equations indicate that the difference of $J$ between two refrigeration regimes is negligibly small for our discussion. Hence, one can focus only on the regime where $V>0$ and $\veps_d>0$ [Figs. \ref{Fig4}(b) and \ref{Fig5}(b)].
Figure \ref{Fig4}(b) displays the detailed map of cooling region with positive bias and the dot level position, where the several contour lines indicate the increasing refrigeration powers in steps of 0.001$\Delta_0^2/h$. One can rediscover the optimum condition for maximum cooling at $eV\approx0.3\Delta_0$ and $\veps_d\approx1.5\Delta_0$ as found in Figs. \ref{Fig2} and \ref{Fig3}. 
In Fig. \ref{Fig4}(c), we show $J$ as a function of $\veps_d$ for various voltages $eV$ selected from horizontal cross sections of Fig. \ref{Fig4}(b). Here, one can more clearly notice an optimum cooling condition.
If the dot level detuning is small, cooling effects are absent regardless of the applied voltage since the the magnitude of particle-hole symmetry breaking does not suffice to induce Peltier refrigeration. At lower voltages, minimum cooling requires a smaller detuning. For $eV=0.1\Delta_0$, this condition takes place at $\veps_d\approx0.6\Delta_0$ from which refrigeration effects are maintained for a broad range of $\veps_d$. In contrast, for $eV=0.5\Delta_0$, the detuning requires $\veps_d\approx1.6\Delta_0$ for cooling effects to start. In addition, the maximum cooling power in this case is rather small and quickly drops down to the heating regime as  Joule heating becomes dominant. At an optimum voltage condition, here for $eV=0.3\Delta_0$, refrigeration starts at a moderate value of the quantum dot level $\veps_d\approx0.8\Delta_0$ and sharply increases up to a maximum value at $\veps_d\approx1.5\Delta_0$. As the dot level is more detuned, cooling effects still persist for a great variety of $\veps_d$ values. Thus, our device does not require fine tunings of the parameters to observe the desired refrigeration effects.
Figure \ref{Fig4}(d) represents $J$ versus $eV$ for several $\veps_d$, i.e., vertical cross sections of Fig. \ref{Fig4}(b), which again confirms the optimum refrigeration condition that we have just described above. Quantitatively, this optimum cooling condition generates the heat flow of $J\sim0.01\Delta_0^2/h$. For $\veps_d<0.5\Delta_0$, strongly nonlinear curves sharply generate large Joule heating effects even at low voltage biases. The curves become strikingly different as $\veps_d$ enhances.

Figure \ref{Fig5} shows the results generated at the elevated temperature $T=0.9T_c$, in comparison to Fig. \ref{Fig4} with $T=0.5T_c$. As expected from Fig. \ref{Fig3}, a higher temperature below $T_c$ tends to generate stronger cooling power. We here find $J\sim0.05\Delta_0^2/h$ with $\veps_d\sim1.5\Delta_0$ and $eV\sim0.3\Delta_0$ [Figs. \ref{Fig5}(c) and \ref{Fig5}(d)].
In Fig. \ref{Fig5}(b), the parameter space for refrigeration effects is larger than that with $T=0.5T_c$, cf. Fig. \ref{Fig4}(b). Indeed, as shown in Fig. \ref{Fig5}(c), a lower value of $\veps_d$ can break the particle-hole symmetry enough to generate the Peltier cooling. For example, $\veps_d\approx0.3\Delta_0$ can already cool the normal metal with $eV=0.1\Delta_0$. The cooling effect persists for a broad range of $\veps_d$ and this range becomes narrower with higher voltages while increasing the maximum power. Here, the increase in refrigeration effect with higher temperatures is due to the suppressed Andreev Joule heating $J_A$ and the thermally activated quasiparticles which positively contributes to the energy current $J_Q^E$. Indeed, in Fig. \ref{Fig5}(d), even for a relatively small detuning of dot level $\veps_d=0.5\Delta_0$, Peltier refrigeration occurs in stark contrast to the case shown in Fig. \ref{Fig4}(d).

\begin{figure}[t]
\centering
    \includegraphics[width=0.4\textwidth]{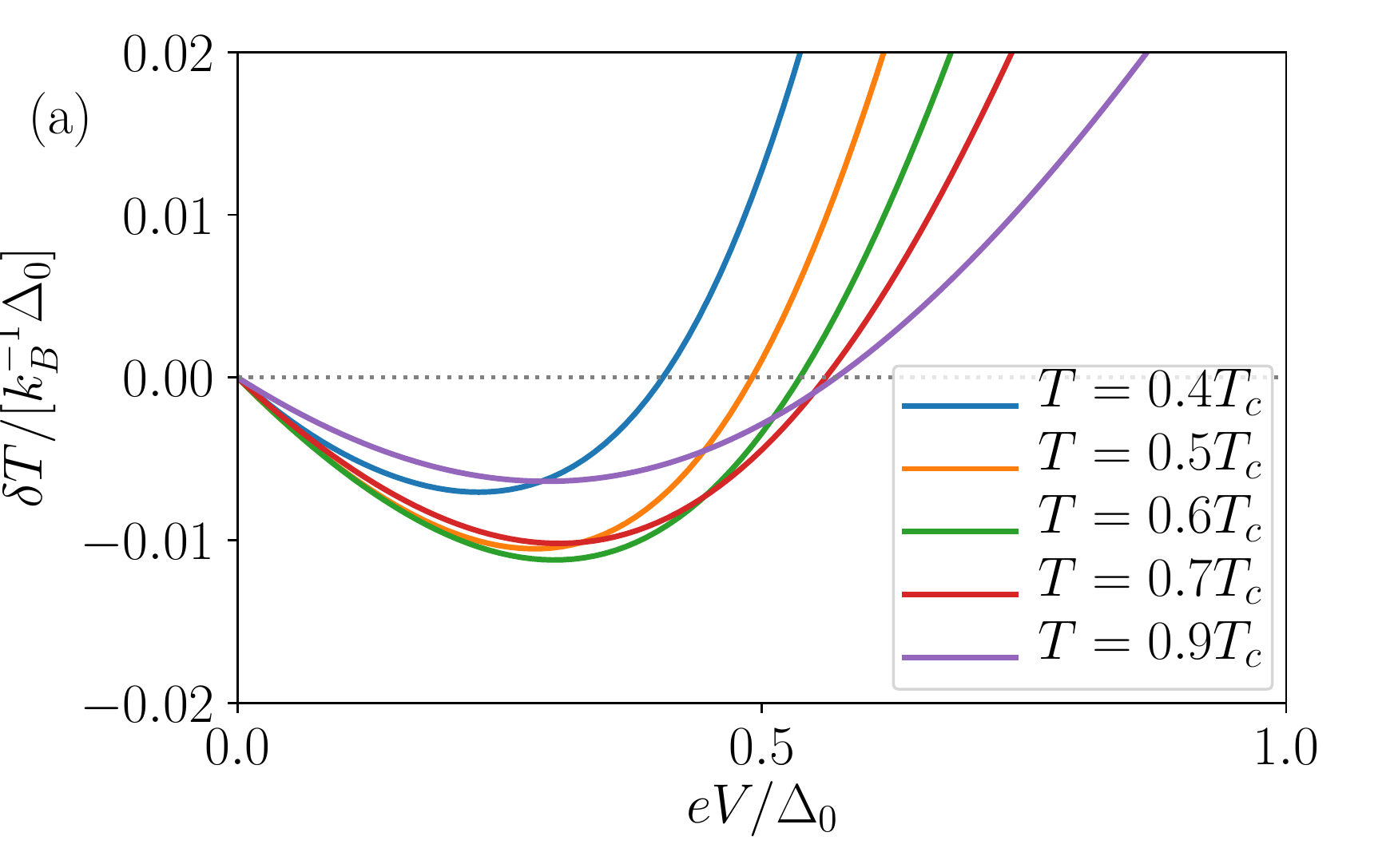}\\
 \includegraphics[width=0.4\textwidth]{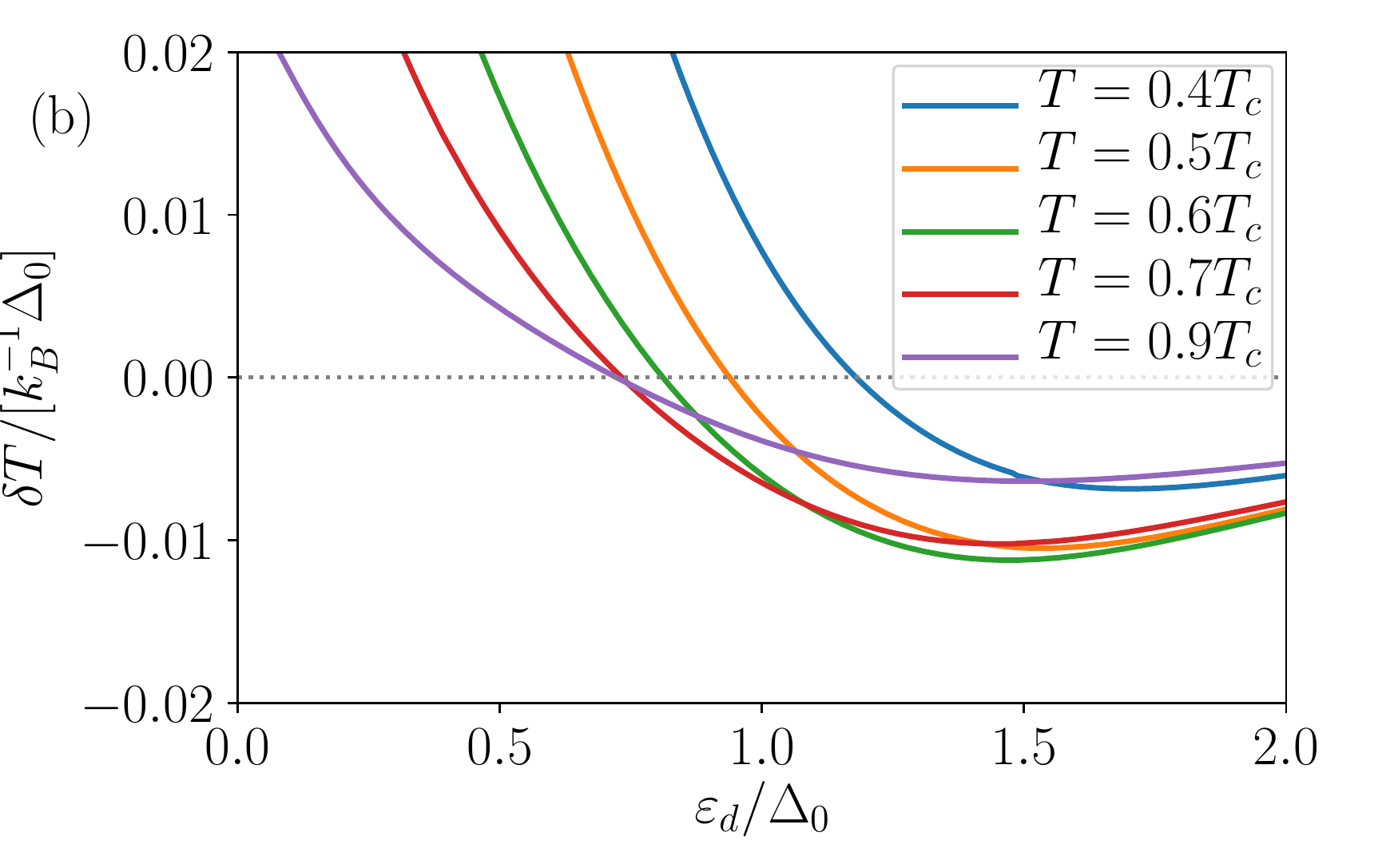}
  \caption{$\delta T$ versus (a) $eV$ at $\veps_d=1.5\Delta_0$, (b) $\veps_d$ at $eV=0.3\Delta_0$. Other parameters are $\Sigma=10^9$~WK$^{-5}$m$^{-3}$, $\mathcal{V}=10^{-20}$m$^{3}$, cf. Eq. \eqref{eq:el-ph}, $\Gamma_L=\Gamma_R=0.5\Delta_0$. }
  \label{Fig6}
\end{figure}

We next estimate the stationary temperature drop of the cooled reservoir due to Peltier effect. Figure \ref{Fig6}(a) displays $\delta T$ as a function of the applied voltage $V$ for several temperatures above $0.4T_c$, below which only the heating dominates [see Fig. \ref{Fig3}(a)]. The quantum dot level is fixed at $\veps_d=1.5\Delta_0$. Interestingly, the temperature drop is optimal around $T=0.6T_c$ although the cooling power $J$ becomes larger as the temperature is increased as we have thus far explained, cf. Fig. \ref{Fig3}. If we take aluminum (Al) as our superconductor, i.e., $\Delta_0\simeq0.34$ meV with $T_c\simeq1.2$ K, we find a large temperature drop around $\delta T\simeq$40 mK at the base temperature $T\simeq600$ mK carried by the cooling power $J\simeq50$ fW, and $\delta T\simeq$20 mK at the base temperature $T\simeq900$ mK with $J\simeq300$ fW. The required voltage to achieve this is of the order of 0.1 mV.
For Niobium (Nb) supercondutors with $\Delta_0\simeq3.05$ meV and $T_c\simeq9.26$ K, the estimations are $\delta T\simeq$355 mK at  $T\simeq4.5$ K with $J\simeq4$ pW, and $\delta T\simeq$180 mK at  $T\simeq8$ K with $J\simeq20$ pW, where the voltage bias $V\simeq0.9$ mV should be applied to the normal metal.
 We also plot in Fig. \ref{Fig6}(b) $\delta T$ versus $\veps_d$ at a fixed voltage bias $eV=0.3\Delta_0$. As we have explained previously, for level detunings $\veps_d\to0$ there is only heating, i.e. $\delta T>0$, even if we sweep over the voltage bias to find its optimum value. Both in Figs. \ref{Fig6}(a) and \ref{Fig6}(b), the temperature drop $\delta T$ becomes small as $T\to T_c$ irrespective of the cooling power strength, cf. Fig. \ref{Fig3}.

In what follows, we characterize the refrigeration efficiency of our device by introducing the coefficient of performance for the cooling defined as the ratio between the cooling power and the input electrical power,
\beq\label{eq:COP}
\text{COP}=\frac{J}{IV}\,,
\edq
which is bounded from above by the Carnot efficiency of refrigeration
\beq\label{eq:carnot}
\veps_C=-\frac{T+\delta T}{\delta T}\,.
\edq
These definitions are valid for $J>0$ in Eq. \eqref{eq:COP} and $\delta T<0$ in Eq. \eqref{eq:carnot} since cooling efficiency is meaningful only in the refrigeration regime ($J>0$) with the corresponding temperature drop ($\delta T<0$).

\begin{figure}[t]
\centering
    \includegraphics[width=0.4\textwidth]{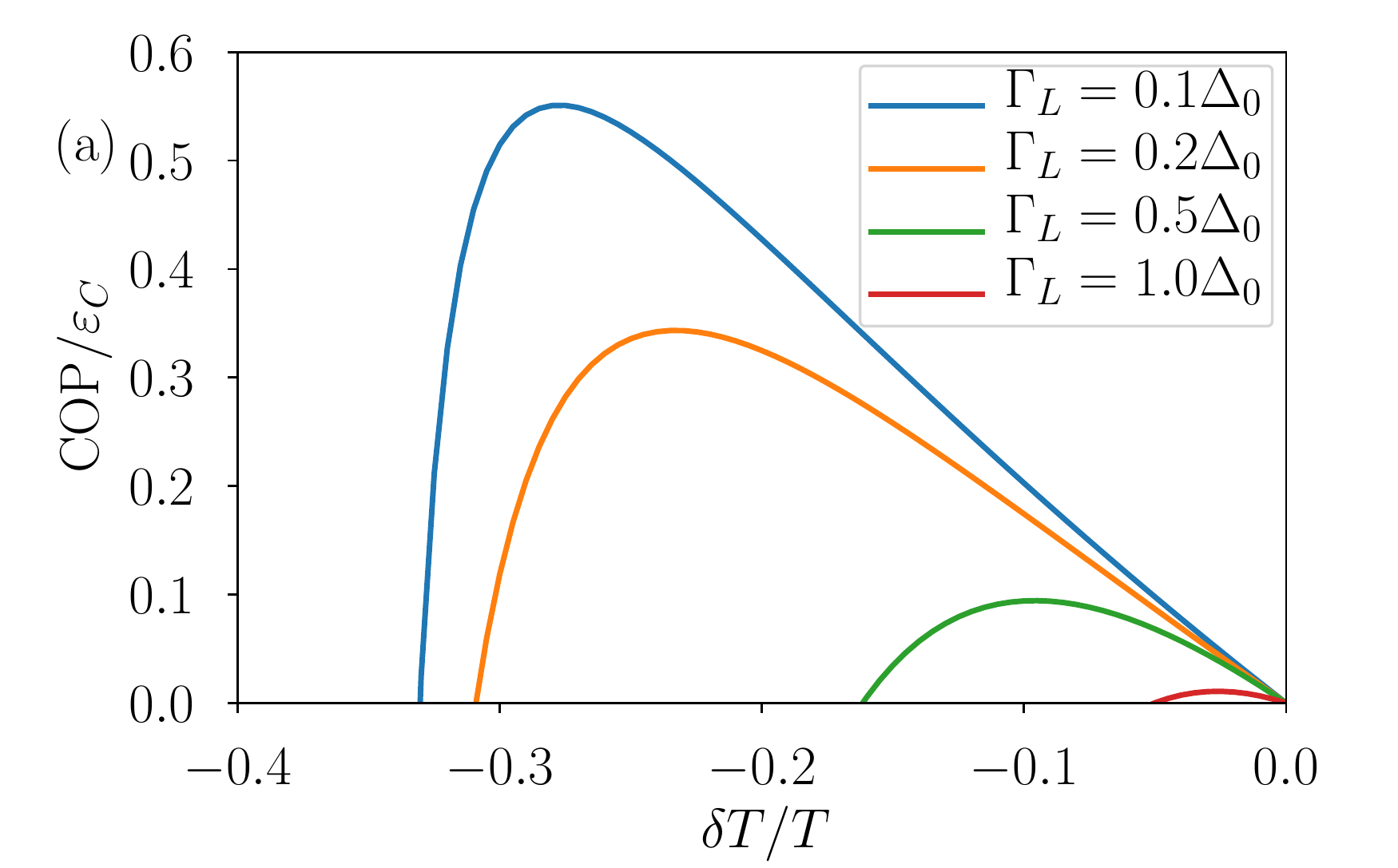}\\
 \includegraphics[width=0.4\textwidth]{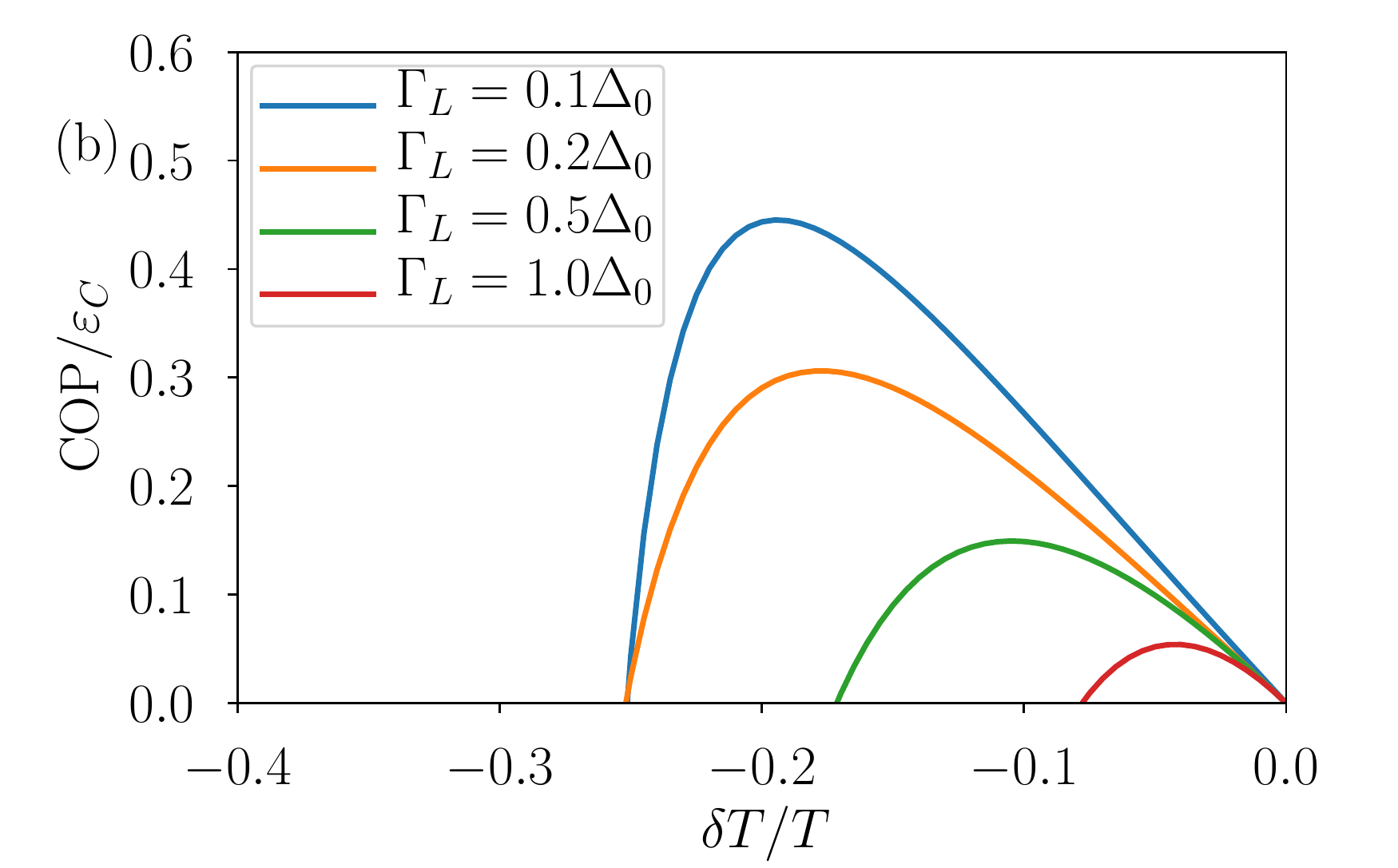}
  \caption{COP vs $\delta T$ for (a) $T=0.5T_c$, (b) $T=0.9T_c$ for various $\Gamma_L=\Gamma_R$ for optimum gate and voltage conditions.}
  \label{Fig7}
\end{figure}

Figure \ref{Fig7}(a) shows the COP in units of the Carnot efficiency as a function of $\delta T/T$ at $T=0.5T_c$ for varying coupling strength $\Gamma_L=\Gamma_R$. Each graph is taken at an optimum condition of the gate and voltage such that the largest cooling power is generated at $\delta T=0$. Remarkably, the coefficient of performance can reach a value slightly larger than half of the Carnot efficiency, i.e., $0.5\veps_C$, if the coupling strength is lowered down to $\Gamma_L=\Gamma_R=0.1\Delta_0$. However, it should be noted that there is a trade-off between the cooling power and the efficiency. Indeed the cooling power is proportional to the square of the coupling strength, i.e., $\Gamma_{L/R}^2$, thus if we take the example of Nb (Al) superconductor, the cooling power becomes of the order of 1 pW (15 fW) to obtain 0.5$\veps_C$. Nevertheless, this quantity is readily detectable with the current experimental technology \cite{fornieri_0-pi_2017,timossi_phase-tunable_2018,dutta_thermal_2017}. For $\Gamma_L=\Gamma_R=0.5\Delta_0$, i.e., the coupling parameters considered in this paper, the device can work at 0.1$\veps_C$ at the maximum efficiency. If the temperature is elevated, as shown in Fig. \ref{Fig7}(b) with $T=0.9T_c$, the COP also decreases slightly for weak couplings. Interestingly, however, for stronger coupling examples with $\Gamma_L=\Gamma_R=0.5\Delta_0$ and $\Gamma_L=\Gamma_R=\Delta_0$, the COP is quite increased compared to the case at $T=0.5T_c$ in Fig. \ref{Fig7}(a), and hence providing the condition to enhance both the cooling power and the efficiency as the cooling power becomes larger at higher temperatures.

\section{Conclusions}\label{conclusion}
In summary, we have proposed a powerful mesoscopic refrigerator based on the superconductor-quantum dot hybrid setup. The normal metal attached to this device can be refrigerated by tuning the gate potential applied to the quantum dot without any ferromagnetic components or magnetic fields. This is due to the strongly energy-dependent transport through the quantum dot. We have considered the full nonlinearity in voltage bias by numerically solving the Coulomb potential of the quantum dot self-consistently. The latter electron-electron interactions are determined at the mean-field level, excluding strong correlation effects. At optimum conditions, the refrigeration power of our device can reach $0.05\Delta_0^2/h$ corresponding to hundreds of fW in case of Al superconductor and a few tens of pW for Nb superconductor. The temperature drop can be of the order of $0.01k_B^{-1}\Delta_0$ which is equivalent to a few tens and hundreds of mK for Al and Nb, respectively. Finally, the refrigeration efficiency can be highly boosted up to half of the Carnot efficiency of cooling at the expense of the power being reduced.
Our results are thus relevant within current efforts for the design of miniaturized engines operating at low temperatures.

\begin{acknowledgments}
SYH thanks W. Wnuczy\'nska for encouraging discussions.
We acknowledge financial support from the Ministry of Innovation NRW via the ``Programm zur Förderung der Rückkehr des hochqualifizierten Forschungsnachwuchses aus dem Ausland'' and the MCIN/AEI/10.13039/501100011033 via the Mar\'{\i}a de Maeztu project CEX2021-001164-M
and the Grant No.\ PID2020-117347GB-I00.
\end{acknowledgments}

\bibliographystyle{apsrev4-1}
\bibliography{NDScooler}
\end{document}